\documentclass[12pt,a4paper]{article}
\usepackage[english]{babel}
\usepackage{graphicx}
\usepackage[]{color}
\usepackage{xcolor} 
\usepackage{alltt}
\usepackage{bm}
\usepackage{bbm}
\usepackage{bbold}
\usepackage{amsmath, amsfonts}
\usepackage{natbib}
\usepackage{color, colortbl} 
\usepackage[colorlinks,citecolor=purple!60!black,linkcolor=blue!60!black,urlcolor=purple!60!black]{hyperref}
\usepackage{comment} 

\usepackage{framed}
\usepackage{float} 
\usepackage{lscape} 
\usepackage{booktabs} 
\usepackage{color, colortbl} 

\usepackage{calrsfs}
\DeclareMathAlphabet{\pazocal}{OMS}{zplm}{m}{n}
\newcommand{\La}{\pazocal{L}} 
\newcommand{\Ra}{\pazocal{R}} 

\usepackage{geometry}
\usepackage{tabularx}
\usepackage{anyfontsize} 

\definecolor{Gray}{gray}{0.95}

\makeatletter
\def\maxwidth{ %
  \ifdim\Gin@nat@width>\linewidth
    \linewidth
  \else
    \Gin@nat@width
  \fi
}
\makeatother
\newcolumntype{Y}{>{\centering\arraybackslash}X} 
\makeatletter
 {\par\unskip\endMakeFramed%
 \at@end@of@kframe}
\makeatother

\usepackage[font=small]{caption}

\usepackage{array}
\newcolumntype{P}[1]{>{\centering\arraybackslash}p{#1}} 

\title{What Drives Inflation and How? Evidence from Additive Mixed Models Selected by cAIC\thanks{We thank an anonymous referee, Jan-Egbert Sturm, David R\"ugamer and the participants of the Swiss National Bank Brown Bag Workshop for their helpful comments and suggestions. The views expressed in this paper are those of the author(s) and do not necessarily reflect those of the Swiss National Bank.}}
\author{Philipp F. M. Baumann\thanks{KOF Swiss Economic Institute, Department of Management, Technology, and Economics (D-MTEC), ETH Zurich, 8092 Zurich, Switzerland. e-mail: baumann@kof.ethz.ch. phone: +41 44 633 87 35}
\and Enzo Rossi\thanks{Swiss National Bank, 8022 Zurich, Switzerland and University of Zurich, 8032 Zurich, Switzerland. e-mail: enzo.rossi@snb.ch. phone: +41 44 631 00 00}
\and Alexander Volkmann\thanks{Chair of Statistics, School of Business and Economics, Humboldt University of Berlin, 10099 Berlin, Germany. e-mail: alexander.volkmann@hu-berlin.de. phone: +49 30 2093 99554}}

\date{\today}
\begin{document}

\maketitle
\pagenumbering{roman}
\thispagestyle{empty}


\begin{abstract}
\noindent
We analyze the forces that explain inflation using a panel of 122 countries from 1997 to 2015 with 37 regressors. 98 models motivated by economic theory are compared to a boosting algorithm, non-linearities and structural breaks are considered. We show that the typical estimation methods are likely to lead to fallacious policy conclusions which motivates the use of a new approach that we propose in this paper. The boosting algorithm outperforms theory-based models. We confirm that energy prices are important but what really matters for inflation is their non-linear interplay with energy rents. Demographic developments also make a difference. Globalization and technology, public debt, central bank independence and political characteristics are less relevant. GDP per capita is more relevant than the output gap, credit growth more than M2 growth.
\end{abstract}











\clearpage

\pagenumbering{arabic}


\section{Introduction}

\begin{figure}[H]

    \centering
    \includegraphics[width=\textwidth]{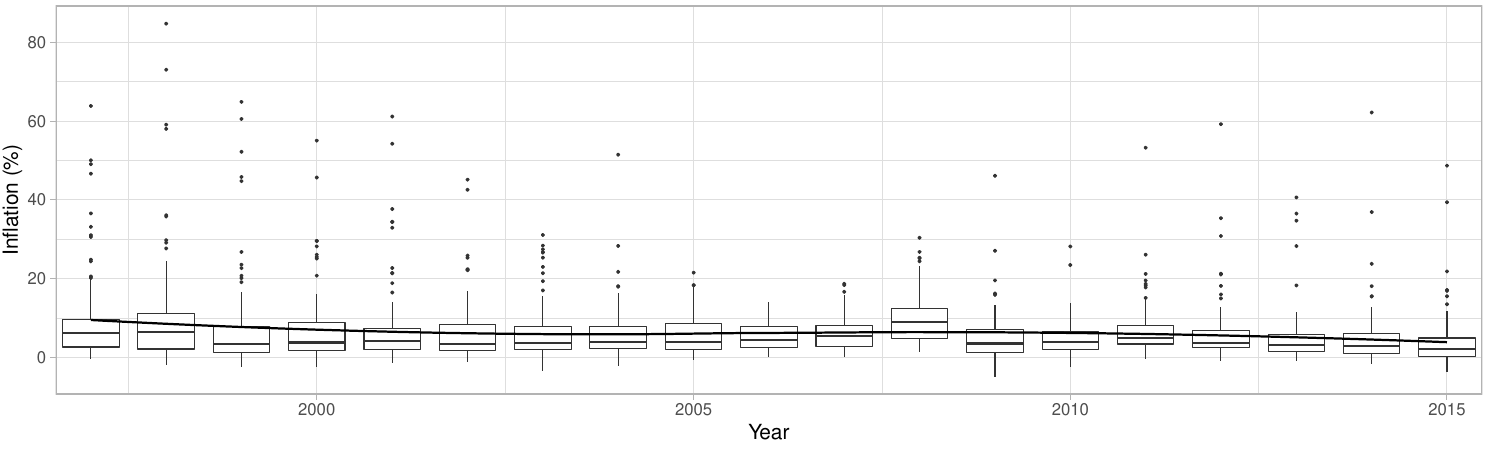}
    \caption[]{Truncated (99.5\% percentile) distribution of inflation over time across 122 countries with a LOESS estimate.}
    \label{fig:BoxPlot}

\end{figure}

In the late 1970s and early 1980s, many countries experienced high inflation. A broad consensus emerged that this was unacceptable. Accordingly, policymakers worldwide adopted or were enabled to adopt policies designed to bring inflation down. As can be inferred from Figure \ref{fig:BoxPlot}, one of the most striking developments of the past two decades has been a steadily declining trend in inflation measured by consumer price index (CPI) and its volatility. In 1997, the average inflation was 21 \%. By 2015 it had dropped to 5 \%.

Many factors are believed to have contributed to this development. They range from stronger commitments to price stability, improved monetary policy, the emergence of the New Economy and the attendant acceleration of productivity growth, forces of globalization that increased competition and enhanced the flexibility of labor and product markets, the weakening influence of trade unions, disciplined fiscal policy, favorable exogenous circumstances, and even luck. All these factors likely played a role, and disentangling the relative contribution of each remains an important challenge.

The general acceptance that the key objective of monetary policy should be price stability has aroused considerable interest in understanding the determinants of inflation. Empirical work in a cross-country setup is broad and diverse in its conclusions. Most of it addresses few potential sources for a limited number of countries or periods. Model comparisons are hardly made, and non-linearities have often not been analyzed. Robustness checks with alternative estimation techniques are rare. Moreover, commonly applied estimation methods are too restrictive and exhibit low explanatory power which in the end may lead to fallacious policy conclusions. We corroborate this hypothesis in Appendix \ref{sec:MotivationAMM} where we show that additive mixed models (AMMs) outperform common estimation techniques.   

Empirical work that takes these shortcomings into account may help improve our understanding of what explains the inflation process over time and across countries. This offers the background to our paper, which identifies and quantifies various determinants of inflation, and motivates our extension of the empirical literature along several dimensions.

First, since the behavior of inflation has become increasingly difficult to understand,\footnote{\cite{Blanchard2016} and \cite{Borio2017} have even put into question economists' knowledge of its process.} we tested several models and variables based on abundant theoretical and empirical research. The explanatory variables were properly lagged to account for potential causal links.

Second, although the downward trend is a global phenomenon that had been noted years ago \citep{Rogoff2003} research has typically focused on low-inflation (advanced) countries. For this reason, we base our analysis on not only as many theoretical explanations as possible but also the highest number of countries, including advanced countries, emerging market and developing economies (EMDEs), as well as low-income countries (LICs). To this end, we pre-processed and analyzed an exceptionally large and comprehensive data set, including annual observations of 37 explanatory variables for 122 countries during the period from 1997 to 2015.

Third, to properly consider the longitudinal structure of the data and the countries' heterogeneities, we recurred to mixed models whose variables were motivated by economic theory on the one hand and by a data-driven variable selection procedure on the other hand. Next, to allowing for a combination of countries with different characteristics, we extended the literature -- which is focused on linear regressions, where inflation is regressed against a specific variable and control variables -- by accounting for potential non-linear relationships between inflation and the regressors. For this purpose, we introduce additive mixed models to empirical research on inflation (which may find application in macroeconomic analyses in general) and provide the software implementation of conditional Akaike Information Criteria (cAIC) for additive mixed models with observation weights. The resulting assignment of cAIC to the models enabled us to compare several theories and the data-driven approach to one another.

The remainder of this paper is organized as follows. In Section \ref{sec:lit} we review the literature and the ensuing explanatory variables underlying our empirical models. Section \ref{sec:Data} presents the data. In Section \ref{sec:Methods} we lay out our estimation methods and the model selection procedure. The main findings are summarized in Section \ref{sec:results}. Section \ref{sec:Conclusion} concludes the paper. Data and code for the reproduction of the analysis is available on \href{https://github.com/PFMB/InflAMM}{GitHub}.

\section{Literature}\label{sec:lit}


A number of empirical studies show that the sources of inflation are quite diverse and include excess demand or slack, a country's institutional set-up, the monetary policy strategy in place, fiscal imbalances, globalization and technology, demography, (shocks to) prices of natural resources, and past inflation. We discuss them in more detail, explain the choice of variables and their abbrevations for the empirical analysis in Appendix \ref{sec:ResultsLiterature}. For an exhaustive survey of the literature we refer to the working paper version of this study \citep{Baumann.2021b_AMM}.

Based on this literature survey, we set up eight economic theories and various testable models, which capture a diversity of country characteristics. In the literature, money and output-related variables are often part of the explanatory variables. For this reason, they are also included in each of our models. Because it is not straightforward which variables best reflect the development of money stock and GDP, each model includes either \textit{M2 Gr. (\%)} or \textit{Credit (\% GDP) Gr. (\%)} in combination with either \textit{GDP pc (USD)} or \textit{GDP Gr. (\%)}, extended by theory-specific explanatory variables. As a result, we obtained a range of four to 24 alternative specifications. This gave rise to an estimation of 98 model-specific variable combinations. The exact variable combinations of each economic model can be gleaned from the Appendix \ref{sec:ResultsAppendix}. In addition to variable compositions suggested by economic theory, we also predefined interactions of variables for which we assumed the existence of a mutual impact on inflation. This applies to \textit{En. Prices (USD) and En. Rents (\% GDP)} as well as \textit{Trade Open. (\% GDP) and Fin. Open}.

\section{Data}\label{sec:Data}

Our aim is to cover as many countries as possible. This entails a trade-off between the number of countries and the completeness of the data set. We were able to collect annual data running from 1995 to 2015 for 124 countries for 21 explanatory variables and for the dependent variable from publicly accessible sources, mainly the IMF and the World Bank. For inflation we finally relied on the IMF's change in the CPI due to data availability. Further, we derived growth rates from level variables, rolling averages from growth rate variables and further transformations from level variables. 30 variables and the dependent variable resulted from this with missing information for some variables (2.8\% of the observations). We imputed the missing observations by means of an EM-Algorithm on bootstrapped samples \citep{Amelia}. We limited the analysis to a single imputation instead of multiple imputed data sets due to the lack of theoretical background for averaging random effects. The contemporaneous measurements of the resulting variables were replaced by their one- or two-year lagged counterpart according to theory and empirical results.\footnote{See \cite{Baumann.2021_causal} for details.} We excluded two countries with outliers from our sample since these countries heavily impaired model selection. This led to 122 countries spanning from 1997 to 2015. We refer to these data as the full sample. In addition to the 30 variables, we also collected eight explanatory variables from various scientific publications and the World Bank that were not available across the whole time span from 1995 to 2015 or were only available for a subset of countries. Due to a non-compliance with the Missing At Random assumption these predictors were not imputed. These variables are associated with the economic theories \textit{Institutions, Monetary Policy Strategies, Public Finance} and \textit{Globalization and Technology}. Their limited availability is one of the reasons for our two-stage selection procedure described in Section \ref{sec:MethodsSelection}.

Finally, this gives rise to a classic longitudinal/panel data structure for a data set comprising 37 predictors and the World Bank's income classification. We provide summary statistics in the Appendix (cf. Figure \ref{fig:Numerics} and \ref{fig:Cats}). According to the World Bank's income classification, approximately 21\% of the countries are low-income countries, 35\% belong to the lower-middle-income category, 19\% to the upper-middle-income category, and 25\% to the high-income category.

\section{Methods}\label{sec:Methods}

In this section we discuss the details of the statistical models and procedures underlying the analysis. First, we present the basic structure of AMMs on which we rely to model annual inflation rates. To capture the country-specific correlation and the heterogeneity of countries, we specify these AMMs with either subject-specific random or fixed effects and country-specific weights. All estimated AMMs are compared by their cAIC. We discuss the cAICs' central pillars in the context of AMMs and present our contribution to its software implementation in Section \ref{sec:cAIC} in the Appendix. We provide the details to model-based boosting for variable selection which we used as the starting point of our data-driven inflation modeling. We then present the two-stage model selection procedure that we developed. Finally, we add varying coefficients based on \citet{Hastie.1993} to the AMMs that exhibit the lowest cAIC to tackle the question of a structural break during the financial crisis 2007/2008.

\subsection{Additive Mixed Models}\label{sec:MethodsAMM}

Mixed models are a natural choice for modeling longitudinal data and have been frequently applied, for example, in epidemiology.\footnote{See, e.g., \citet{degruttola1991modeling} and \citet{pearson1994mixed}.} However, to our knowledge, mixed models have not been applied to model inflation. In general, mixed models include (population) fixed and (subject-specific) random effects. When modeling macroeconomic data, a violation of the random effects assumption may arise, which eventually leads to inconsistent estimators \citep{Wooldridge.2010}. For this reason, we apply a procedure proposed by \citet{Mundlak.1978} to check if the random effects assumption holds or if random effects have to be replaced by country-specific fixed effects. In general, the country-specific effects should act as surrogates for effects that have not been measured and induce heterogeneity between countries. Further, since non-linear relationships between the many predictors used in this paper and inflation cannot be excluded, we extend the mixed models in an additive manner by model terms which are functional forms of the predictors. This leads to the class of AMMs on which our main analysis is based.

The formal structure of the AMMs is as follows: 37 (metric and categorical) predictors and the dependent variable \textit{inflation} (in percent), denoted by $\Tilde{y}_{i,t}$, are given for $i = 1,\ldots, n=122$ countries and for $t = 0,\ldots, T=18$ consecutive years from 1997 to 2015 such that $\bm{\Tilde{y}_{i}} = (\Tilde{y}_{i,0},\ldots, \Tilde{y}_{i,T})^\top $. The vector $\bm{\Tilde{y}} = (\bm{\Tilde{y}_{1}}^\top,\ldots,\bm{\Tilde{y}_{n}^\top)^\top}$ has been transformed by the natural logarithm $\bm{y} := \ln(\bm{\Tilde{y}} + 10.86)$ after shifting the support to values $ \geq 1$ to avoid numerical instabilities. We chose the natural logarithm transformation to meet the distributional assumptions specified for $\bm{\epsilon_{i}}$ in \eqref{eq:AMM}. The generic AMM used to explain $y_{i,t}$ by a set of predictors $A_{j,l}$ is given in Equation \eqref{eq:AMM}. Each of the eight economic theories is represented by a set $G_l:=\{\{A_{1,l}\},\{A_{2,l}\},\ldots,\{A_{m_{l},l}\}\}, l = 1,\ldots,8$, containing $m_l:=\ \mid G_l\mid$ sets of predictors $A_{j,l}$. Each $A_{j,l}$ is composed of disjunct subsets $B_{j,l}$ and $C_{j,l}$ of predictors with linear and non-linear effects, respectively, as well as pairs $D_{j,l}$ of variables in $B_{j,l}$ and pairs $E_{j,l}$ of variables in $C_{j,l}$ with linear and non-linear interaction effects, respectively. Non-linear effects $h$ of predictors $x \in C_{j,l}$ are estimated by univariate cubic P-splines \citep{PSplines} with second-order difference penalties. Interaction effects $f(\cdot,\cdot)$ of pairs $(x, x^{*})$ of variables in $E_{j,l}$ are modeled using penalized bivariate tensor-product splines. The assignment to $B_{j,l}$, $C_{j,l}$, $D_{j,l}$ and $E_{j,l}$ can be found in Tables \ref{Tab:Pred1} and \ref{Tab:Pred2} in the Appendix. Each model $M_{j,l}$ corresponding to one $A_{j,l} \in G_l$ is of the following form:
\begin{equation}\label{eq:AMM}
y_{i,t} =  \beta_{0} 
+ \eta_{i,t}
+  \bm{Z_{i,t}b_{i}} + \epsilon_{i,t},
\end{equation}
\begin{equation}\label{eq:LinPred}
\eta_{i,t} = \sum_{x \in  B_{j,l}} x_{i,t} \beta_{x} 
+ \sum_{(x,x^{*}) \in D_{j,l}}  (x_{i,t}x_{i,t}^{*}) \beta_{(x,x^{*})}
+ \sum_{x \in C_{j,l}} h_x(x_{i,t}) 
+ \sum_{(x,x^{*}) \in E_{j,l}}  f_{(x,x^{*})}(x_{i,t},x_{i,t}^{*})  
\end{equation}
with $\bm{b_{i}} = (b_{i,0},b_{i,1})^\top \stackrel{iid}{\sim} N(\bm{0},\bm{G})$, where a random intercept $b_{i,0}$ and a random slope $b_{i,1}$ with design vector $\bm{Z_{i,t}}\equiv \bm{Z_{t}} = (1,t)$ and non-diagonal covariance $\bm{G}$ are (always) included to capture the serial within-country correlation. Further, $\bm{\epsilon_{i}} \sim N(\bm{0},\bm{R_{i}})$ is assumed with $\bm{\epsilon_{i}} \perp \!\!\! \perp  \bm{b_{i}}$, where $\bm{R_{i}}$ is a diagonal matrix with potentially heterogeneous country-specific variances $\sigma_{i}^{2}$ on its diagonal. The observation weights $w_{i} = \sigma^2/\sigma_{i}^{2}$ emerge implicitly and are contained on the diagonal of the matrix $\bm{\Tilde{W}_{i}}$ such that $\bm{R_{i}} = \sigma^{2} \bm{\Tilde{W}_{i}^{-1}}$. On a sample level, the error covariance structure is a block-diagonal matrix $\bm{R}$ with $\bm{R_{i}}$ on its diagonal.

Assuming $\bm{\epsilon_{i}} \perp \!\!\! \perp  \bm{b_{i}}$ further implies that the $\bm{b_{i}}$ have to be uncorrelated with all $x_{i,t}$ and $x_{i,t}^{*}$ included in the subsets of $A_{j,l}$ for all $t$. This assumption may seem unreasonable given the data and question under investigation. For this reason, we question the random effects assumption (i.e., $\mathbb{E}[\bm{b_i}|x_{i,t},x_{i,t}^{*}] = \mathbb{E}[\bm{b_i}]$ for all $t$) and alternatively specified country-specific fixed effects to take the country-specific correlations into account. Specifically, we alternatively specified each $M_{j,l}$ as in \eqref{eq:AMM} but without any distributional assumption for the country-specific parameters:
\begin{equation}\label{eq:FixedEffect}
    y_{i,t} =
\eta_{i,t}
+  \bm{Z_{i,t}\gamma_{i}} + \epsilon_{i,t},
\end{equation}
where $\bm{\gamma}_{i} = (\gamma_{i,0},\gamma_{i,1})^\top$. To decide if the random effects assumption under \eqref{eq:AMM} or the country-specific fixed effects under \eqref{eq:FixedEffect} are more reasonable for each $M_{j,l}$, we follow \citet{Mundlak.1978}, whose procedure enables us to derive a statistical test which examines if the time-invariant error components of the error in \eqref{eq:AMM} might not be correlated with the time-varying regressors specified in \eqref{eq:LinPred}. We test this hypothesis by specifying a further model $\Bar{M}_{j,l}$ for each $M_{j,l}$ which is specified as the corresponding $M_{j,l}$ but with a linear predictor $\Bar{\eta}_{i,t}$ that additionally encloses the time-averaged transformations of the regressors (i.e., $\Bar{x}_i = \frac{1}{T + 1}\sum_{t = 0}^{T}x_{it}$) 
specified in \eqref{eq:LinPred}. As a result, each $\Bar{M}_{j,l}$ is of the form
\begin{equation}\label{eq:MundlakAMM}
y_{i,t} =  \beta_{0} 
+ \Bar{\eta}_{i,t}
+  \bm{Z_{i,t}b_{i}} + \epsilon_{i,t},
\end{equation}
\begin{equation}\label{eq:MundlakLinPred}
    \Bar{\eta}_{i,t} = \eta_{i,t} + \sum_{x \in  B_{j,l}} \Bar{x}_{i} \Bar{\beta}_{x_B} 
+ \sum_{(x,x^{*}) \in D_{j,l}}  \Bar{x}_{i} \Bar{\beta}_{x_D} + \Bar{x}_{i}^{*}  \Bar{\beta}_{x^{*}_D}
+ \sum_{x \in C_{j,l}} \Bar{x}_{i} \Bar{\beta}_{x_C}
+ \sum_{(x,x^{*}) \in E_{j,l}} \Bar{x}_{i} \Bar{\beta}_{x_E} + \Bar{x}_{i}^{*}  \Bar{\beta}_{x^{*}_E} 
\end{equation}

We test if all the parameters of the time-averaged transformations of the regressors are jointly zero, i.e., $H_0:  \Bar{\beta}_{x_B} = \Bar{\beta}_{x_D} = \Bar{\beta}_{x^{*}_D} = \Bar{\beta}_{x_C} = \Bar{\beta}_{x_E} = \Bar{\beta}_{x^{*}_E} = 0$ against the alternative $H_A$, that at least one of these parameters differs from zero based on a likelihood-ratio test (cf. \ref{sec:MundlakAppendix} in the Appendix). When we could not reject $H_0$ at the 5\%-significance level, we favored the specification under \eqref{eq:AMM} over \eqref{eq:FixedEffect} for the corresponding $M_{j,l}$. We interpret the test result as an indication rather than a statement that provides definitive certainty for the choice between fixed and random effects.

In total, there are 98 (= $\sum_{l = 1}^{8} m_{l}$) such AMMs for all predictor sets $A_{j,l}$ associated with each economic theory $G_l$. For each $G_l$ there is one set of models $\mathcal{M}_l$ which includes all corresponding $M_{j,l}$. We estimated these AMMs by (penalized) maximum likelihood with the \texttt{mgcv} package \citep{mgcv2011} and the \texttt{gamm4} package \citep{wood2016generalized} as extensions to the statistical software \texttt{R} \citep{RSoftware}.

\subsection{Model-based Boosting}\label{sec:MethodsBoosting}
In order to maximize predictive performance on out-of-sample data, we in addition relied on a machine learning approach which could also be used for forecasting purposes. Specifically, we apply a model-based boosting algorithm \citep{Buehlmann.2007, hofner2014, mboost}. It disregards the block by block segmentation of the predictors presented in Section \ref{sec:lit}, which was based upon the associated economic theories. For this reason, we next want to find an optimal prediction function $f^*$ for $\bm{y}$ through some prediction function $f$ which is found by minimizing the expected loss $\mathbb{E}_{Y, X}[\La(\bm{y}, f(\bm{x}))]$ (i.e. risk) through a gradient descent algorithm in function space \citep{hofner2014}. We assume that $f$ is composed of a sum of functions of predictors and country-specific random effects which are all parameterized through different base learners. We specified 34 base learners. The choice of base learners, the employed loss function, the gradient descent algorithm and the selection procedure are discussed in Section \ref{sec:BoostingAppendix} in the Appendix. Finally, 14 out of the 34 base learners were selected. By stopping the algorithm before it converges, a shrinkage effect is imposed onto the effect estimates of the model. Therefore, we refitted the predictors associated with the 14 base learners collected in the predictor set $A_{B}$ as an AMM as specified in \eqref{eq:FixedEffect} and dubbed it $M_{B}$. We favored \eqref{eq:FixedEffect} over \eqref{eq:AMM} due to the result of the testing procedure proposed by \citet{Mundlak.1978}.

\subsection{Selection Procedure}\label{sec:MethodsSelection}

The model selection procedure is as follows: At a first stage $\mathcal{S}_{fir}$, a winner model $M^{*}_{l}$ with the lowest cAIC \eqref{eq:cAICwithR} among models $M_{j,l}$ in the set $\mathcal{M}_l$ is selected for each economic theory. At a second-stage, $M^{*}_{l}$, $l = 1,\ldots, 8$ and $M_{B}$ are collected in the set $\mathcal{M}_{P}$. Some predictors associated with $\mathcal{M}_2$, $\mathcal{M}_3$, $\mathcal{M}_4$ and $\mathcal{M}_5$ are not imputed as these predictors are not available either across time and/or countries which makes a direct model comparison by means of the Likelihood and thus the cAIC invalid. As a result, if the predictor sets included in $M^{*}_{2}$, $M^{*}_{3}$, $M^{*}_{4}$ and $M^{*}_{5}$ are only available for a subsample of data, they are instead added to $\mathcal{M}^{''}$ to be compared to the AMM with the lowest cAIC in $\mathcal{M}_{P}$ later. 
The winner $M_P$ has the lowest cAIC in the set of models $\mathcal{M}_{P}$ and its cAIC is finally compared to each $M^{''}\in \mathcal{M}^{''}$ on the corresponding different data subsets to yield the overall winner $M^{**}$. If the computation of any AMM on any subset of the data fails, this AMM is assigned the highest cAIC in the given comparison. This can happen in particular for complex models on smaller subsets of the data. First- and second-stage selection are together labeled $\mathcal{S}_{sec}$. $M^{**}$ represents the model with the highest empirical relevance and provides the most reasonable set of inflation drivers.  

The reasoning behind this two-stage approach is twofold. First, from a monetary economics perspective it is not known a priori which set of predictors has the most explanatory power for each economic theory ($G_l$). Second, the availability of certain predictor sets $A_{j,l}$ across time and countries enforces this procedure to ensure an admissible model comparison by means of the Likelihood and thus cAIC.

\section{Results}\label{sec:results}

The results of the AMMs presented in Section \ref{sec:Methods} are discussed in this section which is organized in two subsections. In the first, we present the results of $\mathcal{S}_{fir}$ as described in Section \ref{sec:MethodsSelection}. Ordered by theory, we present the winning models, $M^{*}_l$, assessed by their cAIC and the resulting variables, discuss the linear links and plot the pattern of the variables that were estimated as P-splines together with their pointwise 95\% confidence intervals \citep{Wood.2013}. The empirical degree of non-linearity is assessed based on the effective degrees of freedom (EDFs; \citet{Wood2017a}) associated with each penalty specified in \eqref{eq:LinPred}. The EDFs are reported along the y-axis. For example, an EDF equal to 1 indicates that the estimated $M_{j,l}$ penalized the corresponding smooth term to a linear relationship. To solve the identifiability issue of the AMMs specified in Section \ref{sec:MethodsAMM}, all splines estimated incorporate a sum-to-zero constraint (e.g., $\sum_{i,t}\hat{h}_{GDP pc (USD)}(GDP pc (USD)_{i,t}) = 0$ for $M_{6,1}$). As a result, the corresponding effects can only be interpreted on a relative scale. In addition, for each model term enclosed by either \eqref{eq:LinPred} or \eqref{eq:VaryCoef} exhibited in the Appendix \ref{sec:AppendixMethods} we performed a statistical test \citep{Wood.2013}, where under the null the parameters associated with this model term are equal to zero. The order of magnitude of the p-value associated with this test is reported by means of asterisks.\footnote{When $x$ corresponds to the EDF or the linear effect, $x^{***}$ corresponds to significance at the $0.1\%$ level, $x^{**}$ at the $1\%$ level, $x^{*}$ at the $5\%$ level, $x^{\cdot}$ at the $10\%$ level and no asterisks indicates no significance at the $10\%$ level.} Simultaneously, we evaluate the existence of structural breaks in the wake of the financial crisis and juxtapose the evidence of the pre-crisis period with that after the crisis. To this end we applied the varying coefficient approach as defined in Section \ref{sec:VarCoef} in the Appendix. As discussed in Section \ref{sec:Data} and \ref{sec:MethodsSelection}, not all models could be estimated and compared on the full sample. The models included in $\mathcal{M}_1$, $\mathcal{M}_2$, $\mathcal{M}_6$, $\mathcal{M}_7$ and $\mathcal{M}_8$ were fitted on the maximum of observations possible.

For the estimates of institutional characteristics, $\mathcal{M}_2$, we fitted on 26 countries and for a time span from 2000 to 2012 at the first stage. However, we refitted $M^{*}_{2}$ on the full sample at the second stage since the predictors attached to its predictor set, $A_{16,2}$, are available for all 122 countries and all 19 points in time. 
The models examining monetary policy strategy variables, $\mathcal{M}_3$, were fitted on 30 countries and a time interval from 1997 to 2012. The models examining effects from public finance, $\mathcal{M}_4$, were fitted on 79 countries from 1997 to 2015. The AMMs enclosed by $\mathcal{M}_5$, that is globalization and technology, were fitted on 93 countries and from 1997 to 2012. Since the predictors, $A_{3,3}$, $A_{14,4}$ and $A_{20,5}$, are not available in the full sample, $M^{*}_{3},M^{*}_{4}$ and $M^{*}_{5}$ were excluded from $\mathcal{S}_{sec}$. 

In the second subsection, we describe the results of $\mathcal{S}_{sec}$ which were characterized by the addition of $M_{B}$ to the winners of $\mathcal{S}_{fir}$. Here, we identify the overall winning model, $M^{**}$, and describe its links to log inflation. 

\subsection{First Stage Selection}\label{sec:FirstStageSelection}

This subsection describes the results organized by economic theory. It first presents the winning model within the estimated model combinations and compares the empirical relevance of the variables involved. The winning model is characterized by the lowest cAIC value. Table \ref{Tab:Pred1} and \ref{Tab:Pred2} in the Appendix display the results divided by theory. The first section lists the results for money, credit and slack, $\mathcal{M}_1$, the second section those for institutions $\mathcal{M}_2$, the third for monetary policy strategies, $\mathcal{M}_3$, the fourth for public finance $\mathcal{M}_4$, the fifth for globalization and technology, $\mathcal{M}_5$, the sixth for demography, $\mathcal{M}_6$, the seventh for natural resources, $\mathcal{M}_7$, and the eighth for past inflation, $\mathcal{M}_8$. As can be gleaned from the p-values reported in Table \ref{Tab:Pred1} and \ref{Tab:Pred2}, we specified country-specific fixed effects rather than random effects if the tested hypothesis specified based on the proposal of \citet{Mundlak.1978} in Section \ref{sec:MethodsAMM} was significant at the 5\% level.

The subsequent discussion focuses on whether the effects are linear or not, based on the EDF values. Three figures plot the results for three different periods of observations. A left-hand panel plots the results for the whole sample, a middle panel those relating to the pre-crisis and a right-hand panel those for the post-crisis period. 

\subsubsection{Money, Credit, and Slack}

\begin{figure}[H]

    \centering
    \includegraphics[width=\textwidth]{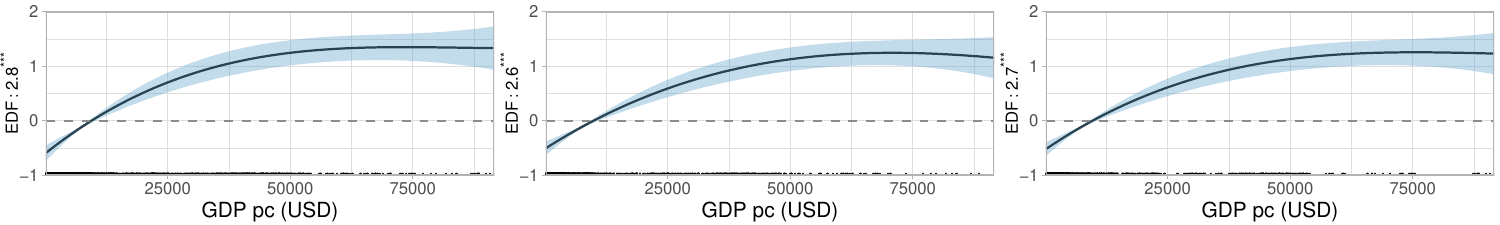}
    \caption[]{The estimate $\hat{h}_{GDP\ pc\ (USD)}(GDP\ pc\ (USD))$ results from the winning model $M_{6,1}$ specified under \eqref{eq:LinPred} and \eqref{eq:VaryCoef}. The corresponding EDFs are reported along the y-axis. The ticks on the x-axis indicate the ranges of strong (dense ticks) and weak (sparse ticks) data support of the GDP pc (USD) variable.}
    \label{fig:M1}

\end{figure}

AMMs that include M2 growth exhibit higher empirical relevance than those that include credit growth, while the models that include output gap are more relevant than those that account for GDP growth. However, the output gap is less relevant than GDP pc. 
$M_{6,1}$ is the winning model. It exhibits GDP pc and credit growth. There is evidence of a linear and positive association between credit growth and log inflation. A one percentage point increase leads to a rise in log inflation of $0.022^{*}$ (approximately 2.2 percent increase in inflation). The estimated effect after the global financial crisis (GFC) ($0.136^{**}$) has strengthened relative to the pre-crisis period ($0.0178^{}$). In contrast, GDP pc affects inflation in a non-linear way as seen in Figure \ref{fig:M1}. An increase up to 50,000 USD is associated with a sharp increase in log inflation and peters out at this income value. 

\subsubsection{Institutions}

\begin{figure}[H]

    \centering
    \includegraphics[width=\textwidth]{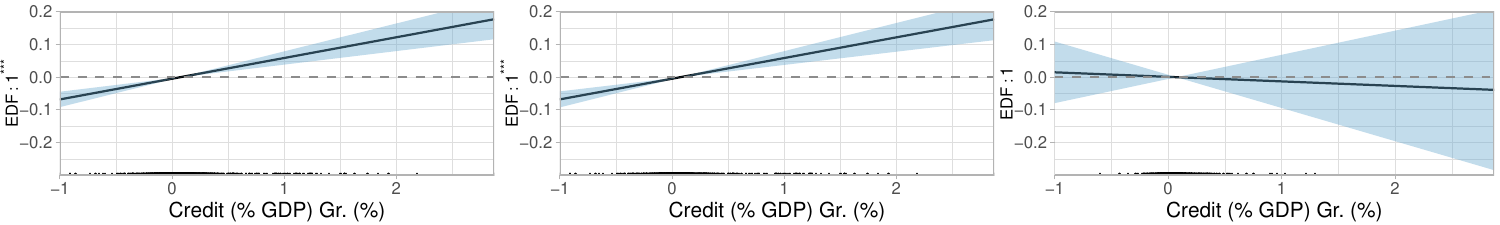}
    \caption[]{The estimate $\hat{h}_{Credit\ (\% GDP)\ Gr. (\%)}(Credit\ (\% GDP)\ Gr. (\%))$ results from the winning model $M_{16,2}$ specified under \eqref{eq:LinPred} and \eqref{eq:VaryCoef}. The corresponding EDFs are reported along the y-axis. The ticks on the x-axis indicate the ranges of strong (dense ticks) and weak (sparse ticks) data support of the Credit\ (\% GDP)\ Gr. (\%) variable.}
    \label{fig:M2}

\end{figure}

In all cases models with credit growth are more relevant than the 
models with M2 growth. GDP growth does better than GDP pc in 10 out of 12 cases. The freedom status variable bears also empirical relevance. However, these results are derived from  a reduced sample size of 26 countries and a period from 2000 to 2012. 
The winning model is $M_{16,2}$ which features civil liberties next to credit growth and GDP pc. Due to the full-sample availability of civil liberties, we refitted the winning model on the full sample. The results are as follows: In the winning model all variables show evidence of a weak linear relationship with log inflation. In particular, credit growth (Figure \ref{fig:M2}) affects log inflation in a linear way. However, after the GFC, as indicated by missing asterisks of the EDFs, it cannot be told if the effect differs from zero. Estimated across the entire time span, the transition from no civil liberties to higher civil liberties is associated with an increasing impact on log inflation ($0.01^{*}$ at most across levels). However, before the crisis, this effect was positive ($0.12$ at most) and turned negative afterwards ($-0.2^{*}$ at most). GDP pc exerts a significant negative effect ($-0.00001^{***}$) across the entire period.

\subsubsection{Monetary Policy Strategies}

\begin{figure}[H]

    \centering
    \includegraphics[width=\textwidth]{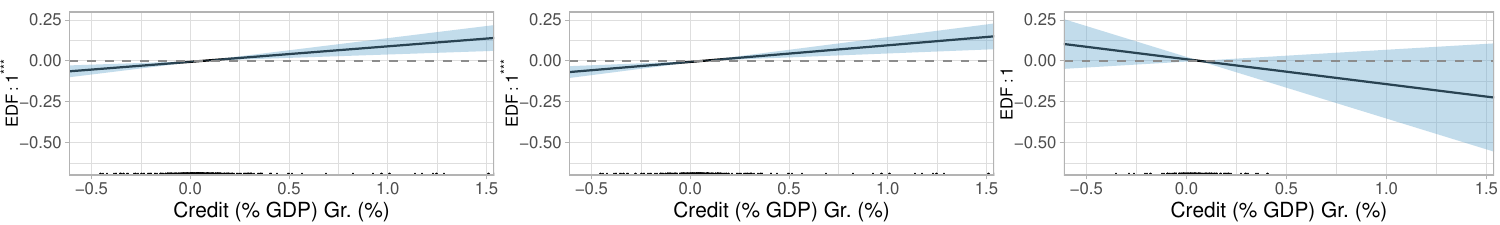}
    \caption[]{The estimate $\hat{h}_{Credit\ (\% GDP)\ Gr. (\%)}(Credit\ (\% GDP)\ Gr. (\%))$ results from the winning model $M_{3,3}$ specified under \eqref{eq:LinPred} and \eqref{eq:VaryCoef}. The corresponding EDFs are reported along the y-axis. The ticks on the x-axis indicate the ranges of strong (dense ticks) and weak (sparse ticks) data support of the Credit (\% GDP) Gr. (\%) variable.}
    \label{fig:M3}

\end{figure}

Models including exchange-rate arrangements (ERA) do better than those with inflation targeting. Credit growth and M2 growth do equally well in terms of empirical relevance, whereas GDP growth outperforms GDP pc in four out of four cases. 
$M_{3,3}$ is the winning model. According to it, ERA are important next to credit and GDP growth. The transition from a situation with no legal tender, actually a fixed-exchange-rate regime, to managed floating leads to a rise in log inflation ($0.052^{*}$). This effect is slightly weaker ($0.031^{}$) for a transition to a crawling-peg and weakest for the transition to free-floating ($0.002^{}$). No structural changes could be estimated for ERA due to singularities. Credit growth displays a positive linear relationship with log inflation (Figure \ref{fig:M3}). This holds before the crisis but vanishes after that, although estimated with high uncertainty. GDP growth also exhibits a linear effect ($1.059^{***}$), which has slightly strengthened after the crisis ($1.439^{***}$). 

\subsubsection{Public Finance}

\begin{figure}[H]

    \centering
    \includegraphics[width=\textwidth]{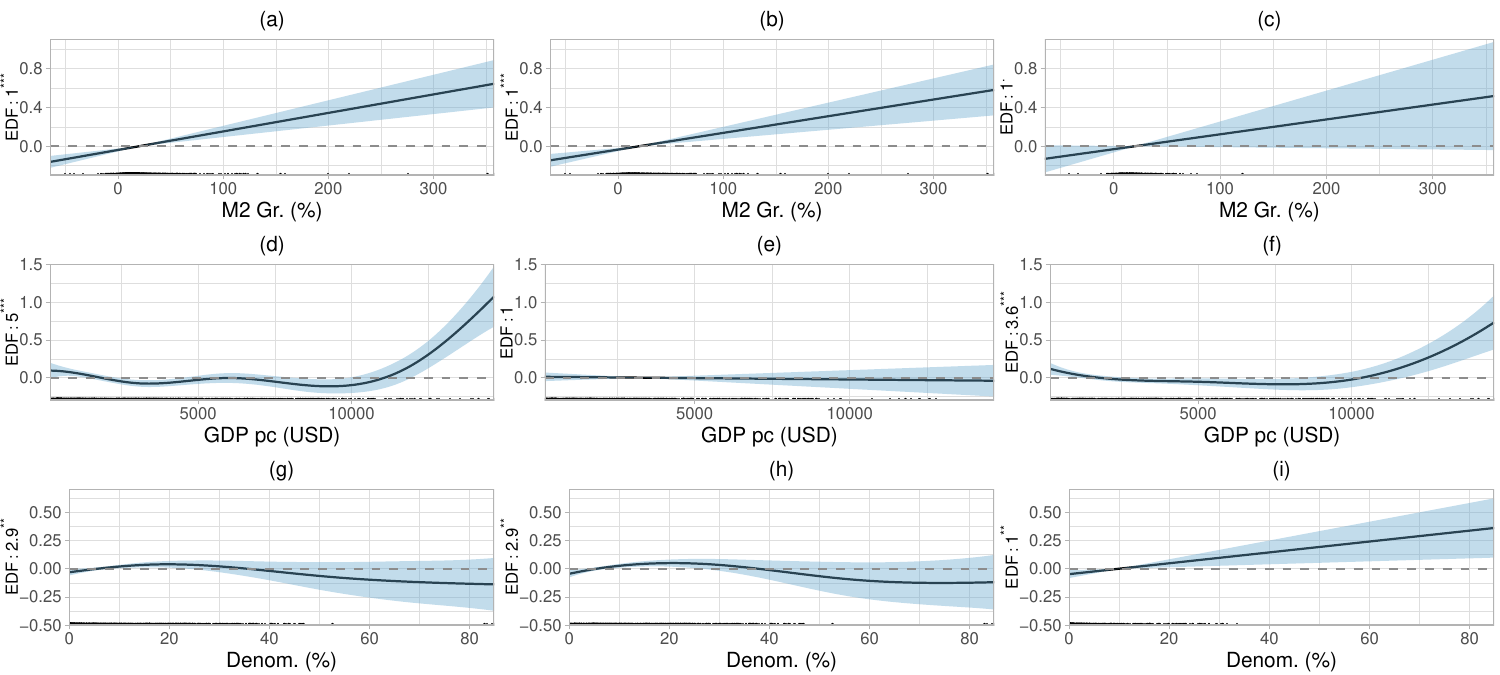}
    \caption[]{The three variables displayed result from the winning model $M_{14,4}$ specified under \eqref{eq:LinPred} and \eqref{eq:VaryCoef}. The corresponding EDFs are reported along the y-axis. The ticks on the x-axis indicate the ranges of strong (dense ticks) and weak (sparse ticks) data support of the variables.}
    \label{fig:M4}

\end{figure}

Models with M2 growth do better than those with credit growth in seven out of eight comparisons. Models with GDP pc turn out to be better than the models that include GDP growth. Debt denomination (Denom. (\%)) plays a dominant role while the maturity structure (Matur.) is less relevant. 
$M_{14,4}$ is the winner. Figure \ref{fig:M4} summarizes the estimations which exhibit some non-linearities. It includes M2 growth, GDP pc and debt denomination. M2 growth exhibits a positive linear link with log inflation. In contrast, GDP pc (USD) reveals a clear non-linear link (panel d). While the effect varies somehow below a threshold of 10,000 USD, it strongly increases beyond this income level. This pattern arises after the crisis (panel f). Debt denomination exhibits a cubic association with log inflation over the entire period (panel g). Beyond a share of public and publicly guaranteed external long-term debt denominated in a foreign currency of 20\%, a further issuance reduces log inflation. The comparison between the pre-crisis period  summarized in panel h with the post-crisis period (panel i), shows a clear break. Since then, increasing the share of foreign-currency debt linearly boosts log inflation. Due to data availability, this evidence is obtained for observations of low and middle-income countries where an effect may be more likely than in advanced countries. However, the results after the crisis contrast with theoretical predictions from the time-inconsistency literature. One possible explanation is that the more debt is issued in a form that protects investors from unexpected inflation, the higher the level of inflation required to reduce the inflation-sensitive part of the debt.

\subsubsection{Globalization and Technology}

\begin{figure}[H]

    \centering
    \includegraphics[width=\textwidth]{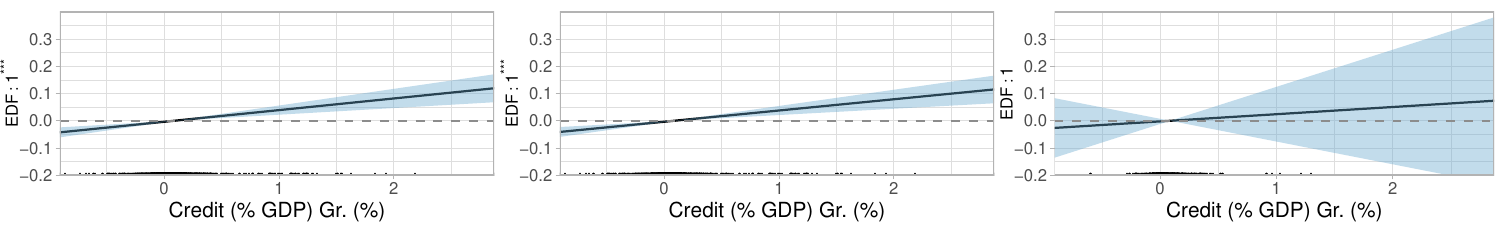}
    \caption[]{The estimate $\hat{h}_{Credit\ (\% GDP)\ Gr. (\%)}(Credit\ (\% GDP)\ Gr. (\%))$ results from the winning model $M_{20,5}$ specified under \eqref{eq:LinPred} and \eqref{eq:VaryCoef}. The corresponding EDFs are reported along the y-axis. The ticks on the x-axis indicate the ranges of strong (dense ticks) and weak (sparse ticks) data support of the Credit (\% GDP) Gr. (\%) variable.}
    \label{fig:5Glob}

\end{figure}

Models with GDP growth are superior to models that exhibit GDP pc  in eight out of nine comparisons. Credit growth stands out in comparison with M2 growth in eight out of eight cases. 
The winning model is  $M_{20,5}$ which features information and communication technology capital over the total capital stock (ICT Capital) next to credit and GDP growth. When ICT Capital is increased by one unit, log inflation rises by $4.088^{***}$ c.p. on average (approximately 4.1 percent increase in inflation). The effect weakens  when separated into the pre-crisis ($2.537^{***}$) and the post-crisis ($2.748^{***}$) era. As illustrated in Figure \ref{fig:5Glob} credit growth reveals a linear link with log inflation over the whole sample period  and in the pre-crisis period but disappears subsequently. In contrast, while GDP growth hardly affected log inflation before the crisis ($0.123^{}$), it has boosted inflation ($1.581^{***}$) thereafter, leading to an inflation rising relationship over the whole period ($0.083^{***}$).

\subsubsection{Demography}

\begin{figure}[H]

    \centering
    \includegraphics[width=\textwidth]{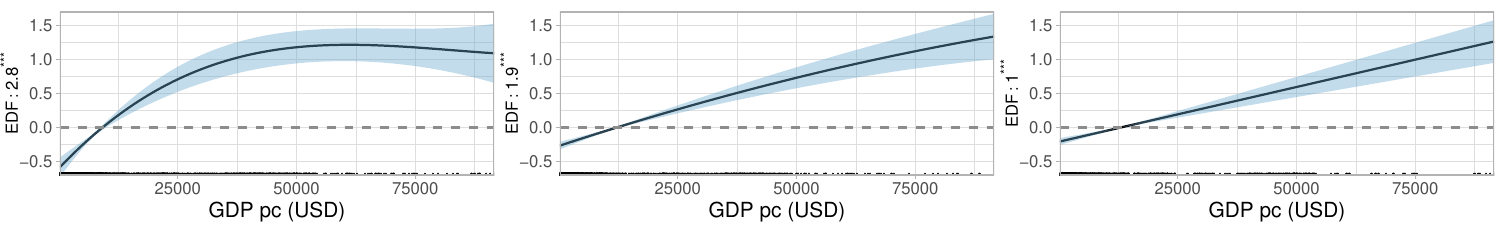}
    \caption[]{The estimate $\hat{h}_{GDP\ pc\ (USD)}(GDP\ pc\ (USD))$ results from the winning model $M_{4,6}$ specified under \eqref{eq:LinPred} and \eqref{eq:VaryCoef}. The corresponding EDFs are reported along the y-axis. The ticks on the x-axis indicate the ranges of strong (dense ticks) and weak (sparse ticks) data support of the GDP pc (USD) variable.}
    \label{fig:6Demo}

\end{figure}

AMMs with credit growth fare better than those with M2 growth in three out of four cases. This also holds for the models featuring the share of the population older than or equal to 65 (Age 65 (\%)) compared to those exhibiting the share of population older than or equal to 75. Models that include GDP pc are superior to the models with GDP growth in three out of four cases. 
$M_{4,6}$ exhibits the variable combination that best explains a relationship between demography and log inflation. It includes Age 65 (\%) next to credit growth and GDP pc. Age 65 (\%) exerts a significant (at all levels) negative effect on log inflation. If this share increases by one percentage point, log inflation decreases on average by $0.039^{***}$. This effect is of similar magnitude before the crisis ($-0.032^{***}$) but weakens afterwards $(-0.022^{***})$. From Figure \ref{fig:6Demo} which displays the non-linear estimates of GDP pc, we can infer a similar non-linearity over the whole sample period as for GDP pc in the winning model $M_{6,1}$ illustrated in Figure \ref{fig:M1}. However, in contrast to $M_{6,1}$, where the non-linearity holds up in all three (sub)periods, the effect changes from quadratic before the crisis to linear after the crisis. Note that we observe higher values of GDP pc after the crisis than before. Credit growth exhibits a positive linear association ($0.023^{*}$) across the whole sample. Before the crisis  the effect is similar to the overall observation ($0.025^{***}$) but strengthens ($0.085^{***}$) after that.

\subsubsection{Natural Resources}

\begin{figure}[H]

    \centering
    \includegraphics[width=\textwidth]{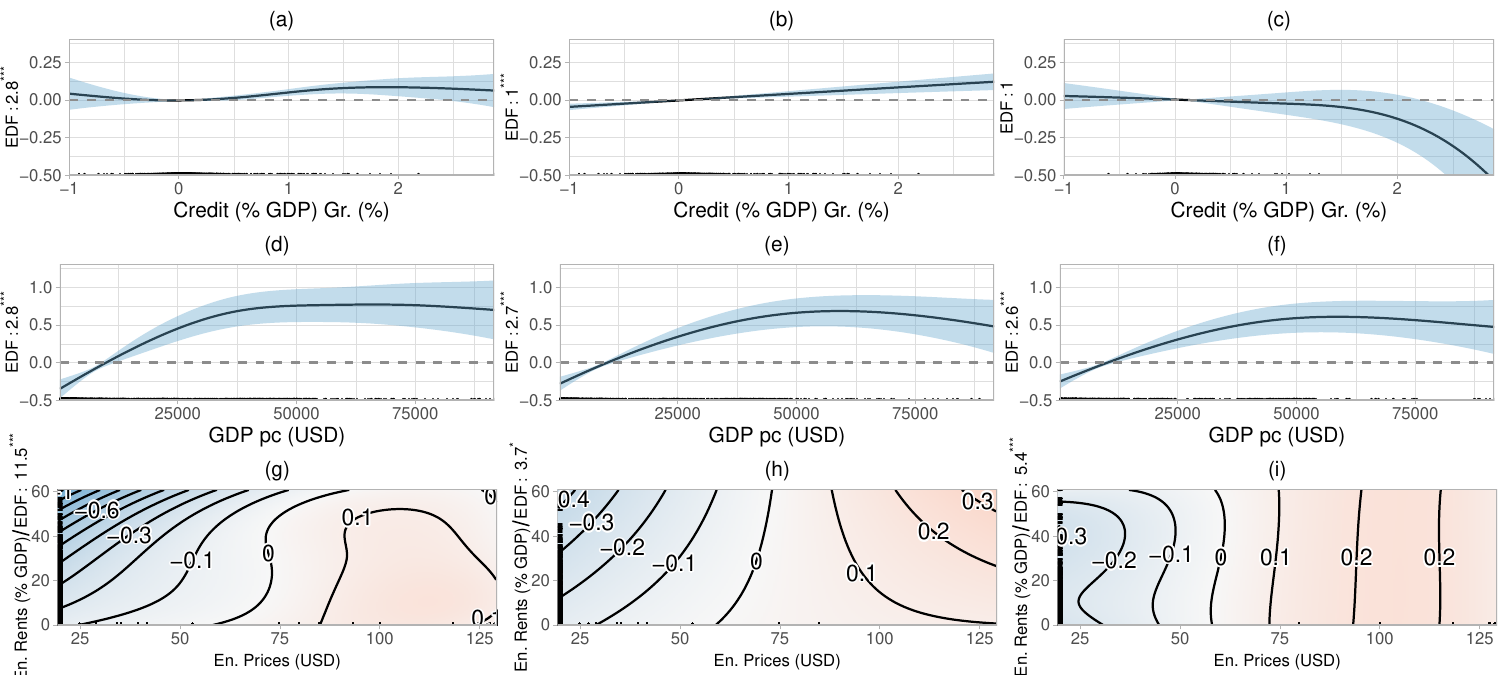}
    \caption[]{The three variables displayed result from the winning model $M_{12,7}$ specified under \eqref{eq:LinPred} and \eqref{eq:VaryCoef}. The corresponding EDFs are reported along the y-axis. The ticks on the x-axis indicate the ranges of strong (dense ticks) and weak (sparse ticks) data support of the variables.}
    \label{fig:NaturalResources}

\end{figure}

In three out of four comparisons, AMMs that include credit growth instead of M2 growth yield a better result. Models with GDP pc are superior to those that contain GDP growth in two out of three cases. 
$M_{12,7}$ results as the winning model. It is composed of credit growth, GDP pc, and the interaction of energy prices with
energy rents. From Figure \ref{fig:NaturalResources} we can infer non-linear relationships. An acceleration in credit growth in the range between 0 and 150\% (panel a) pushes log inflation non-linearly. The effect is positive and linear before the crisis (panel b) but becomes negative and non-linear in the post-crisis period (panel c). Turning to GDP pc (panels d-f), the relationship with log inflation is again similar to Figure \ref{fig:M1}. It is cubic throughout. Panels g-i illustrate the bivariate interaction effects between energy prices  and energy rents using contour plots. They show the joint relationship between energy prices on the x-axis, energy rents on the y-axis, and log inflation. The passage from a blue to a red area denotes mounting inflationary pressure. Conversely, the passage from a red to a blue area indicates a decrease in inflation. The black contour (iso-effect value) lines indicate the strength of the effects which can only be interpreted on a relative scale, as discussed at the beginning of this section. Along the same iso-effect line the interaction effect does not change. From panel g, a non-linear interaction effect between energy prices and rents can be inferred. When energy prices are below 75 USD while rents are high (above 25), the strongest impact on log inflation arises from an increase in energy prices. The effect from rising energy prices beyond 75 USD is still positive but weakens sharply. When energy rents hover below 25, an energy price increase still boosts log inflation, but by a much smaller magnitude than when rents are high. In the pre-crisis period, the relationship remains non-linear (panel h). The post-crisis still exhibits a non-linear interaction as long as energy prices are below 50 USD, but disappears beyond this level. While an increase  in energy prices still leads to more log inflation, a change in energy rents leaves log inflation unaffected for any value of energy prices above 50 USD (c.p.).




\subsubsection{Past Inflation}

\begin{figure}[H]

    \centering
    \includegraphics[width=\textwidth]{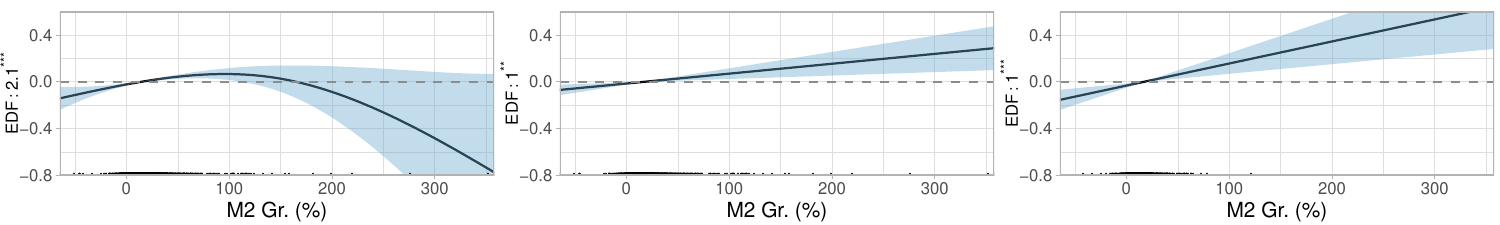}
    \caption[]{The estimate $\hat{h}_{M2\ Gr.\ (\%)}(M2\ Gr.\ (\%))$ results from the winning model $M_{2,8}$ specified under \eqref{eq:LinPred} and \eqref{eq:VaryCoef}. The corresponding EDFs are reported along the y-axis. The ticks on the x-axis indicate the ranges of strong (dense ticks) and weak (sparse ticks) data support of the M2 Gr. (\%) variable.}
    \label{fig:PastInflation}

\end{figure}

AMMs that feature M2 growth strictly outperform AMMs that exhibit credit growth. No clear picture emerges from the comparison of AMMs that include GDP pc with models that include GDP growth. 
The winning model is $M_{2,8}$ and includes past inflation together with M2 growth and GDP pc. There is evidence of a positive linear effect ($0.001^{***}$) from past inflation estimated over the whole sample period. While the relationship did not change, the strength of the effect increased somewhat since the crisis ($0.011^{***}$) compared with the preceding period ($0.001^{***}$). As seen in the left panel of Figure \ref{fig:PastInflation}, there is a quadratic relationship between M2 growth and log inflation in the whole sample, but with evidence for a linear relationship in the region with the most data support. The uneven distribution of the data should limit the interpretation of the effects in areas without any data support. An acceleration of M2 growth below a level of 100\% raises inflation. Beyond 100\%, the impact becomes highly uncertain. In contrast, before the crisis the center panel suggests that M2 Gr.(\%) impacted log inflation linearly. After the crisis, the effect strengthened slightly. In contrast, GDP pc exhibits a positive and linear effect over the whole period ($0.00002^{***}$), before ($0.00002^{***}$) and after the crisis ($0.00002^{***}$).

We do not explicitly examine how people form their inflation expectations. However, the importance of past inflation suggests the existence of (at least a share of) "adaptive expectations users" in practice.





\subsection{Second Stage Selection}

In $\mathcal{S}_{fir}$, discussed in Section \ref{sec:FirstStageSelection}, we derived the winning model for each economic theory. In this subsection, we discuss the derivation of the overall winning model, $M^{**}$. This required a second stage selection because the winning model of the first stage for four of the theories examined was obtained from a lower number of countries and a reduced period. This applies to theories associated with institutions, monetary policy strategies, public finance, and globalization and technology ($\mathcal{M}_{2}$, $\mathcal{M}_{3}$,$\mathcal{M}_{4}$ and $\mathcal{M}_{5}$). Their Likelihood and thus their cAICs cannot be directly compared with the AMMs from the other theories -- $\mathcal{M}_{1}$, $\mathcal{M}_{6}$, $\mathcal{M}_{7}$, $\mathcal{M}_{8}$ -- and the AMM selected by the boosting algorithm, $M_{B}$. However, since some AMMs comprised by $\mathcal{M}_{2}$, $\mathcal{M}_{3}$, $\mathcal{M}_{4}$ and $\mathcal{M}_{5}$ contain $M_{j,l}$ associated with predictor sets $A_{j,l}$ that are also available for the full sample, these $M_{j,l}$ can be refitted on the full sample, in case they were selected during $\mathcal{S}_{fir}$. This is the case for $M^{*}_{2}$. As a result, $M^{*}_{2}$ has been refitted on the full sample and was added to the comparison of the AMMs that  were already estimated during the first stage comparison (i.e. $M_{6,1}, M_{20,5}, M_{4,6}$ and $M_{12,7}$). We next present $M_{B}$ and compare its cAIC against the cAIC of the first-stage winners.

\paragraph{AMM Selected by the Boosting Algorithm}
The boosting algorithm selected the set of predictors $A_B$ which can be inferred from the last subsection of Table \ref{Tab:Pred2} exhibited in the Appendix. This subsection also includes separating all selected predictors into disjoint subsets, informing which predictors were modeled (non-)linearly and/or through a bivariate interaction. For the boosting algorithm we added more predictors to those exhibited in the AMMs presented in Section \ref{sec:FirstStageSelection}. These additional variables are domestic credit level by the financial sector in percent of GDP (Credit Fin. (\% GDP)) and its growth rate (Credit Fin. (\% GDP) Gr. (\%)). The remaining additional variables are M2 (\% GDP), Credit (\% GDP), Debt (\% GDP), En. Price Gr. (\%), En. Rents Gr. (\%), GDP (USD) and GDP pc Gr. (\%).

\begin{figure}[H]

    \centering
    \includegraphics[width=\textwidth]{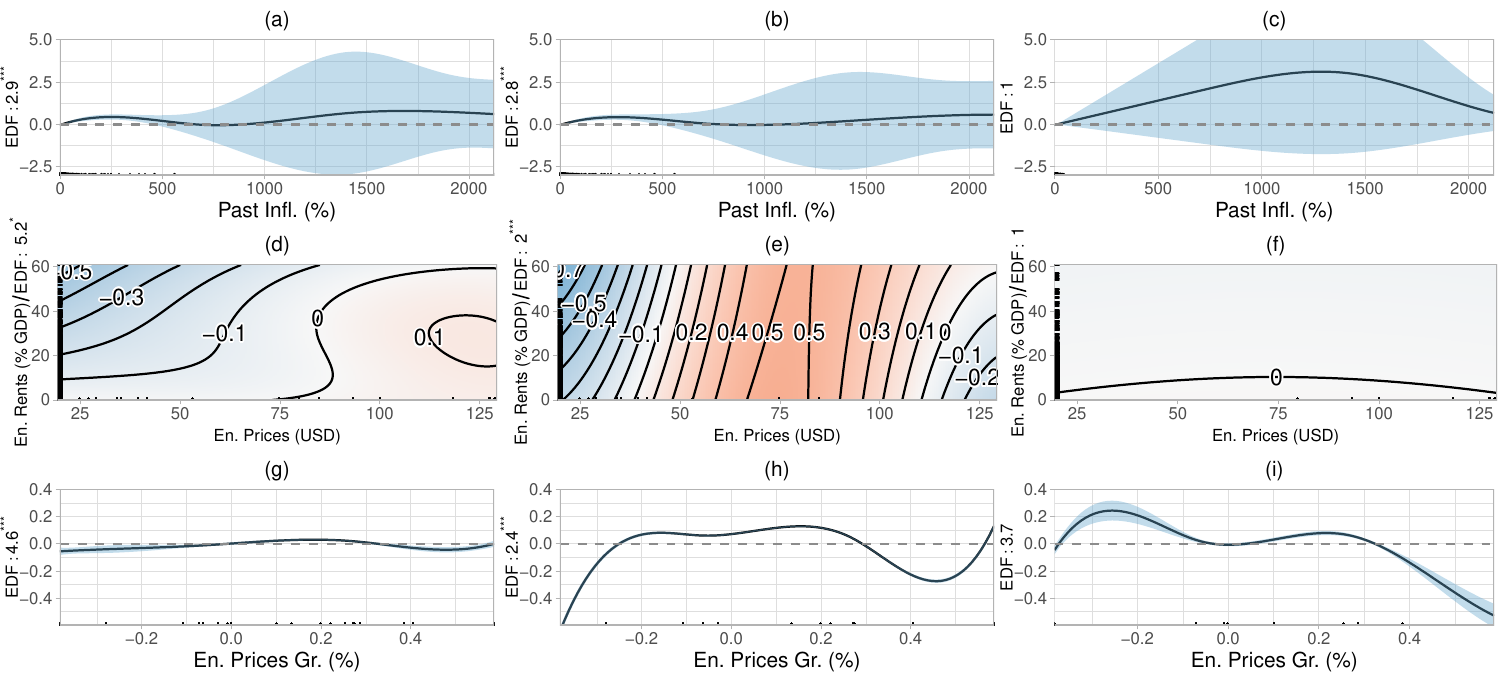}
    \caption[]{The first out of three plots that displays the estimated non-linear effects from $M_{B}$.}
    \label{fig:Boosting1}

\end{figure}

\begin{figure}[H]

    \centering
    \includegraphics[width=\textwidth]{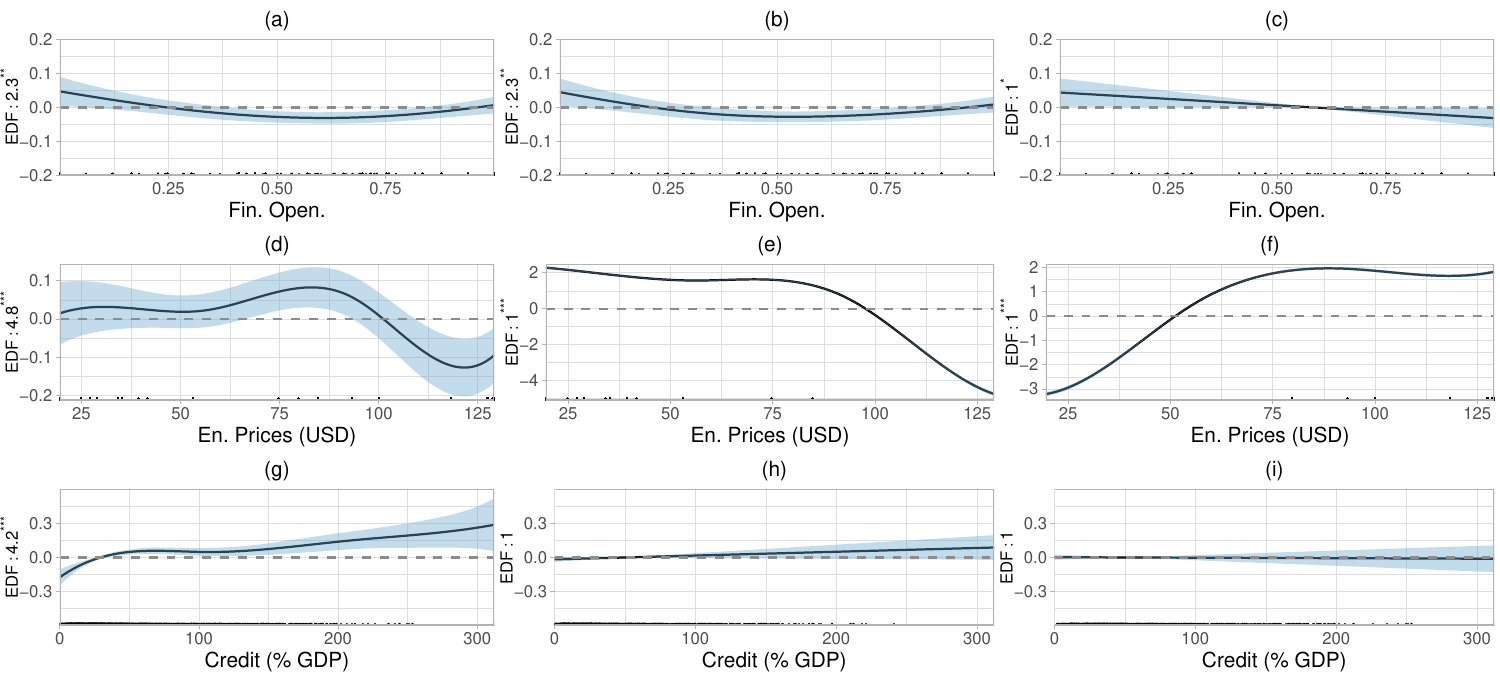}
    \caption[]{The second out of three plots that displays the estimated non-linear effects from $M_{B}$.}
    \label{fig:Boosting2}

\end{figure}

\begin{figure}[H]

    \centering
    \includegraphics[width=\textwidth]{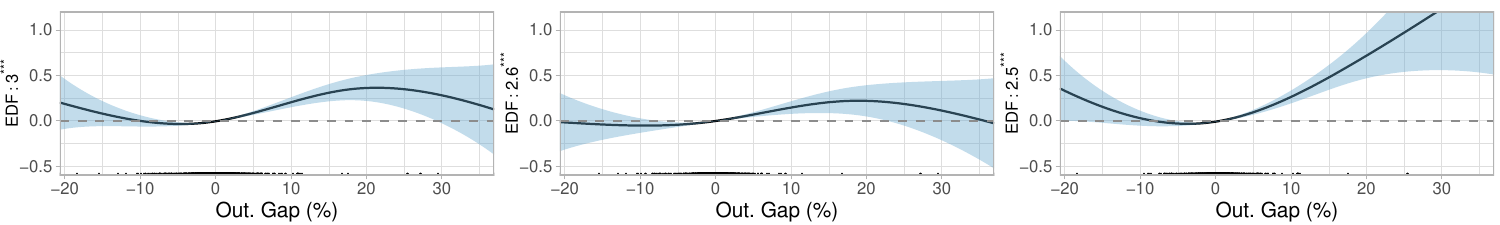}
    \caption[]{The third out of three plots that displays the estimated non-linear effects from $M_{B}$.}
    \label{fig:Boosting3}

\end{figure}

Figure \ref{fig:Boosting1}, \ref{fig:Boosting2} and \ref{fig:Boosting3} present the non-linear effect estimates included in $M_{B}$. Past inflation (Figure \ref{fig:Boosting1}, panel a-c) suggests such a pattern across the whole sample but with high uncertainty when assessed over the whole sample period. However, in the range where most observations lie ($<250\%$), the relationship is linear with a positive slope. The same observation holds for the pre- and post-crisis period. The bivariate interaction of energy prices and rents (panel d) confirms the results from the estimation of $M_{12;7}$ illustrated in Figure \ref{fig:NaturalResources}, at least over the entire sample. In the pre-crisis period the interplay between energy prices and rents weakens (panel e). An increase in energy prices beyond $75$ USD would lower log inflation, irrespective of the value of energy rents. Below energy prices of $75$ USD a rise in energy prices would increase log inflation and still for any level of energy rents. On the other hand, if energy rents rise, there is hardly any effect on log inflation, regardless of the level of energy prices. Finally, in the post-crisis era (panel f), the impact of the interaction of energy prices and rents vanishes completely. When interpreting these results it has to be kept in mind that $M_{B}$ estimates the univariate effects of energy prices and rents in contrast to $M_{12,7}$ which estimates their interaction.

Panels g, h, and i display the results of energy price growth. Over the whole sample, the relationship is highly non-linear (panel g). For a growth rate of energy prices below 20\% a rise in the growth rate increases log inflation. Beyond a growth rate of 20\% a further energy price rise has an inflation abating effect, followed again by an acceleration above 50\%. The evidence for the pre-crisis period can be seen in panel h and in the post-crisis period in panel i. Note that the variation for the energy price variable (and its growth) results exclusively from the time variation and not from the cross-country variation, as we have identical energy prices for each country in the estimation. The resulting uneven distribution of the data weakens the reliability of the effects in areas without any data support.

Figure \ref{fig:Boosting2} summarizes the results of financial openness, energy prices and credit. As long as values of financial openness hover below 0.6, an increase in openness lowers log inflation but increases it beyond this threshold (panel a). This pattern also holds before the crisis (panel b) but turns linear (panel c) after the crisis. Energy prices show again a strong non-linear relationship (panel d). Below 80 USD, a rise in energy prices is conducive to inflation, although subject to high uncertainty. Afterward, the effect turns negative. The evidence preceding the crisis (panel e) suggests that energy prices were associated with lower inflation, especially beyond 80 USD. However, this changed dramatically after the crisis (panel f) where energy price boosts below 80 USD are linked with continuously higher log inflation and stagnation afterward. For credit the relationship is also non-linear but positive and strong for values below 50 over the whole sample (panel g). When separating into the pre- (panel h) and the post-crisis period (panel i), the effect does not differ from zero.

The last non-linear effects comprised by $M_{B}$ are shown in Figure \ref{fig:Boosting3} and relate to the output gap whose pattern suggests a cubic relationship with log inflation. Log inflation is boosted by a widening  gap between -5\% and 20\% and followed by a negative effect (although subject to increasing uncertainty). This pattern holds over the whole sample and in the period preceding the crisis. However, after the crisis, the relationship has changed, becoming negative for output gap values below -5\% and positive after that.

Finally, all linearly estimated effects of $M_{B}$ are insignificant at the 10\% level. One exception is M2 growth ($0.0005^{*}$) whose impact weakens before the crisis ($0.0004^{}$) but turns stronger ($0.003^{***}$) after the crisis. The second exception is trade openness which exhibits a negative impact on log inflation ($-0.001^{*}$) across the whole sample but turns insignificant when separated into a pre- and post-crisis effect.

\paragraph{Overall Winner} As seen in Table \ref{Tab:Pred1} and \ref{Tab:Pred2} in the Appendix the comparison among theory-based winning models yields $M_{12,7}$ as the best model. However, the lowest cAIC overall is exhibited by $M_B$. Since both AMMs feature variables associated with natural resources, we conclude that these variables play a key role in the inflation (disinflation) process. In particular, the interaction of prices and rents of natural resources exhibited in $M_{B}$ and $M_{12,7}$ seems to have particularly high explanatory power. The empirical relevance is higher when energy rents and prices interact than when they enter as two separate univariate terms. Moreover, their interacting effects are highly non-linear. The boosting algorithm supports this interaction and shows the importance of energy prices and their growth rate as additional univariate drivers of inflation. Finally, we computed the cAIC for models that do not contain any economic variable at all (that is, they exhibit no effects other than for time and country-level) and found a substantially higher cAIC for these models compared to every other model included in the first and second stage selection. Consequently, it can be inferred that model compilation based on economic theory significantly improves the goodness-of-fit.

Summarizing the evidence of the pairwise comparisons on a meta-level yields that credit growth outperforms M2 growth and GDP pc outperforms GDP growth.

\section{Conclusions}\label{sec:Conclusion}

We contribute to the literature on what determines inflation and how by estimating a large quantity of macro, institutional and political models in a sample of 122 countries at different stages of development from 1997 to 2015.



From  among  the  eight  theories,  the  winning  model  includes  energy prices whose importance has already been highlighted in previous work. However, we find that the most compelling determinant of inflation are not energy prices alone but their interplay with energy rents which exhibits a strong non-linear  association with inflation. The atheoretical boosting algorithm confirms the importance of the interplay of energy prices with energy rents. It outperforms  all theoretically motivated models in terms of explanatory power and  suggests, in line with previous analyses, a particular role for energy prices. Energy rents by themselves do not seem to be as important.

The results have a bearing on monetary policy. The empirical importance of energy prices (and rents) has implications for when and how central banks need to respond to oil price shocks. Another challenge to monetary policy-making arises from the link between past and current inflation in a low-inflation environment. One way to lift inflation has been the stimulus of credit creation. However, there is little evidence that this policy was successful. It cannot be excluded that it even backfired. A promising tool to boost inflation is higher GDP per capita level. This suggests that economic policies geared towards growth should be more promising than monetary policies aimed at enhancing credit growth. Another result relates to the output gap. While in monetary theory and policy it is considered a key variable in the determination of inflation, it plays a minor role compared to GDP per capita. 



\bibliographystyle{rss}
\bibliography{bibliography.bib}


\appendix


\newgeometry{
  left=1cm,
  right=1cm,
  top=0.5cm, 
  bottom=1.5cm
}
\setcounter{page}{1} 
\section{Appendix: Results}\label{sec:ResultsAppendix}

\begin{table}[H]
\centering
\tiny 
\begin{tabularx}{\textwidth}{p{0.7cm}P{3.2cm}P{3.1cm}P{3cm}P{3cm}P{2cm}P{1.2cm}}
\toprule
 & $\bm{B_{j,l}}$ & $\bm{C_{j,l}}$ & $\bm{D_{j,l}}$ & $\bm{E_{j,l}}$ &  \textbf{p-value} & \textbf{cAIC} \\ 
  \midrule
  \rowcolor{Gray}
  $A_{1,1}$ & M2 Gr. (\%) & Out. Gap (\%), Year & & & 1.00 & -1155.98 \\ 
  $A_{2,1}$ & GDP Gr. (\%) & M2 Gr. (\%), Year & & & 1.00 & -1117.81 \\ 
  \rowcolor{Gray}
  $A_{3,1}$ & GDP pc (USD) & M2 Gr. (\%) & & & 0.00 & -1280.26 \\ 
  $A_{4,1}$ & Out. Gap (\%) & Credit (\% GDP) Gr. (\%), Year &  & & 0.92 & -1170.79 \\ 
  \rowcolor{Gray}
  $A_{5,1}$ & GDP Gr. (\%) & Credit (\% GDP) Gr. (\%), Year &  & &1.00 & -1133.39 \\
  $A_{6,1}$ & Credit (\% GDP) Gr. (\%) & GDP pc (USD) & &  &0.00 & \textbf{-1300.80} \\ 
  \midrule
    \rowcolor{Gray}
  $A_{1,2}$ & M2 Gr. (\%), CBT & GDP Gr. (\%) &  & &  0.98 & -393.34 \\ 
  $A_{2,2}$ & M2 Gr. (\%), GDP pc (USD) & CBT &  &  & 0.37 & -391.92 \\ 
        \rowcolor{Gray}
  $A_{3,2}$ & Credit (\% GDP) Gr. (\%), CBT & GDP Gr. (\%) & & & 0.88 & -395.57 \\ 
  $A_{4,2}$ & CBT, GDP pc (USD) & Credit (\% GDP) Gr. (\%), Year & & & 0.16 & -395.85 \\  
        \rowcolor{Gray}
  $A_{5,2}$ & Pol. Orien., Pol. Stab.,  M2 Gr. (\%) & GDP Gr. (\%), Year & Pol. Orien. : Pol. Stab. & & 1.00 & -401.33 \\ 
  $A_{6,2}$ & Pol. Orien., Pol. Stab.,  GDP pc (USD) &  M2 Gr. (\%), Year & Pol. Orien. : Pol. Stab.& & 1.00 &-397.20 \\ 
        \rowcolor{Gray}
  $A_{7,2}$ & Pol. Orien., Pol. Stab.,  Credit (\% GDP) Gr. (\%) & Year, GDP Gr. (\%) & Pol. Orien. : Pol. Stab. &  & 1.00 &-402.67 \\
  $A_{8,2}$ & Pol. Orien., Pol. Stab., GDP pc (USD) & Credit (\% GDP) Gr. (\%), Year & Pol. Orien. : Pol. Stab. &  & 1.00 &-399.27 \\ 
        \rowcolor{Gray}
  $A_{9,2}$  & M2 Gr. (\%), Pol. Rights & GDP Gr. (\%), Year &  & & 1.00 &-399.52 \\ 
  $A_{10,2}$ & Pol. Rights, GDP pc (USD) & M2 Gr. (\%), Year &  &   & 1.00 &-397.78 \\ 
        \rowcolor{Gray}
  $A_{11,2}$ & Pol. Rights, Credit (\% GDP) Gr. (\%) & GDP Gr. (\%), Year &  & & 1.00 &-401.72 \\
  $A_{12,2}$ & Pol. Rights, GDP pc (USD) & Credit (\% GDP) Gr. (\%), Year &  &  & 1.00 &-400.48 \\ 
        \rowcolor{Gray}
  $A_{13,2}$ & Civil Lib., M2 Gr. (\%) & GDP Gr. (\%), Year &  & & 1.00 &-389.19 \\ 
  $A_{14,2}$ & Civil Lib., GDP pc (USD) & M2 Gr. (\%), Year &  & & 1.00 &203.39  \\ 
        \rowcolor{Gray}
  $A_{15,2}$ & Civil Lib., Credit (\% GDP) Gr. (\%) & GDP Gr. (\%), Year &  &  & 1.00 &-393.18 \\ 
  $A_{16,2}$ & Civil Lib., GDP pc (USD) & Credit (\% GDP) Gr. (\%), Year &  &  & 1.00 & \textbf{-413.58}/ -1115.99 \\  
        \rowcolor{Gray}
  $A_{17,2}$ & Fr. Status, M2 Gr. (\%) & GDP Gr. (\%), Year &  & & 1.00 &-405.67  \\ 
  $A_{18,2}$ & Fr. Status, GDP pc (USD) &  M2 Gr. (\%), Year &  &  & 0.99 &-403.87  \\  
        \rowcolor{Gray}
  $A_{19,2}$ & Fr. Status, GDP pc (USD) & Credit (\% GDP) Gr. (\%), Year &  & & 0.75 &-405.45 \\ 
  $A_{20,2}$ & Fr. Status, Credit (\% GDP) Gr. (\%) & GDP Gr. (\%), Year &  & & 1.00 &-407.99 \\ 
        \rowcolor{Gray}
  $A_{21,2}$ & M2 Gr. (\%), CBI, TOR & GDP Gr. (\%), Year & TOR : CBI & & 1.00 &-396.04 \\ 
  $A_{22,2}$ & CBI, TOR, GDP pc (USD) & M2 Gr. (\%), Year & TOR : CBI & & 0.99 &-392.48 \\  
        \rowcolor{Gray}
  $A_{23,2}$ & Credit (\% GDP) Gr. (\%), CBI, TOR & GDP Gr. (\%), Year & TOR : CBI & & 1.00 &-398.46  \\  
  $A_{24,2}$ & CBI, TOR, GDP pc (USD) & Credit (\% GDP) Gr. (\%), Year & TOR : CBI & & 0.53 &-395.28  \\  
        \midrule
        \rowcolor{Gray}
  $A_{1,3}$ & ERA, GDP Gr. (\%) & M2 Gr. (\%), Year & & & 1.00 & -386.56 \\ 
  $A_{2,3}$ & ERA & M2 Gr. (\%), GDP pc (USD), Year & &  & 1.00 & -368.36 \\ 
        \rowcolor{Gray}
  $A_{3,3}$ & ERA, GDP Gr. (\%) & Credit (\% GDP) Gr. (\%), Year & & & 1.00 & \textbf{-390.12} \\
  $A_{4,3}$ & ERA, Credit (\% GDP) Gr. (\%) & GDP pc (USD), Year & & & 1.00 &-367.31 \\ 
        \rowcolor{Gray}
  $A_{5,3}$ & Infl. Targ., GDP Gr. (\%) & M2 Gr. (\%), Year & & & 1.00 &-381.05 \\ 
  $A_{6,3}$ & Infl. Targ. & GDP pc (USD), M2 Gr. (\%), Year & & & 1.00 &-363.80 \\
        \rowcolor{Gray}
  $A_{7,3}$ & Infl. Targ., GDP Gr. (\%) & Credit (\% GDP) Gr. (\%), Year & &  & 0.30 & -384.77  \\ 
  $A_{8,3}$ & Infl. Targ., Credit (\% GDP) Gr. (\%) & GDP pc (USD), Year & & & 0.52 & -363.59 \\ 
        \midrule
  \rowcolor{Gray}
  $A_{1,4}$ & Debt (\% GDP) Gr. (\%) & M2 Gr. (\%), GDP Gr. (\%), Year & & & 1.00 &-127.18 \\ 
  $A_{2,4}$ & Debt (\% GDP) Gr. (\%) & M2 Gr. (\%), GDP pc (USD), Year & & & 1.00 &-176.95 \\ 
      \rowcolor{Gray}
  $A_{3,4}$ & Credit (\% GDP) Gr. (\%), Debt (\% GDP) Gr. (\%) & GDP Gr. (\%), Year & & & 0.81 &-121.38 \\ 
  $A_{4,4}$ & Credit (\% GDP) Gr,  Debt (\% GDP) Gr. (\%) & GDP pc (USD), Year & & & 1.00 &-137.40 \\ 
      \rowcolor{Gray}
  $A_{5,4}$ & M2 Gr. (\%) & Prim. Bal. (\% GDP), GDP Gr. (\%), Year & & & 1.00 &-133.08 \\ 
  $A_{6,4}$ & Prim. Bal. (\% GDP) & M2 Gr. (\%), GDP pc (USD), Year & & & 1.00 &-177.66 \\ 
      \rowcolor{Gray}
  $A_{7,4}$ & Prim. Bal. (\% GDP), Credit (\% GDP) Gr. (\%) & GDP Gr. (\%), Year & & & 1.00 &-119.48 \\ 
  $A_{8,4}$ & Prim. Bal. (\% GDP), Credit (\% GDP) Gr. (\%) & GDP pc (USD), Year & & & 1.00 &-144.61 \\ 
      \rowcolor{Gray}
  $A_{9,4}$ & Matur. & M2 Gr. (\%), GDP Gr. (\%), Year & & & 1.00 &-120.76 \\ 
  $A_{10,4}$ & Matur. & M2 Gr. (\%), GDP pc (USD), Year & & & 1.00 &-176.68 \\  
      \rowcolor{Gray}
  $A_{11,4}$ & Credit (\% GDP) Gr. (\%), Matur. & GDP Gr. (\%), Year &  & & 1.00 &-115.07 \\ 
  $A_{12,4}$ & Credit (\% GDP) Gr. (\%), Matur. & GDP pc (USD), Year &  & & 1.00 &-133.05 \\ 
      \rowcolor{Gray}
  $A_{13,4}$ & & Denom. (\%), M2 Gr. (\%), GDP Gr. (\%), Year & & & 1.00 &-138.57 \\ 
  $A_{14,4}$ & & Denom. (\%), M2 Gr. (\%), GDP pc (USD), Year & & & 1.00 &\textbf{-262.59} \\ 
      \rowcolor{Gray}
  $A_{15,4}$ & Credit (\% GDP) Gr. (\%) & Denom. (\%), GDP Gr. (\%), Year & &  & 1.00&-138.64 \\ 
  $A_{16,4}$ & Credit (\% GDP) Gr. (\%) & Denom. (\%), GDP pc (USD), Year & & & 1.00 &-159.66 \\ 
 \bottomrule
\end{tabularx}
\caption{(1/2) Allocation of the predictor set $A_{j,l}$ of the model $M_{j,l}$ to $B_{j,l}$,$C_{j,l}$,
$D_{j,l}$ and $E_{j,l}$. The cAIC value for the AMM with the lowest cAIC is printed in bold. For $A_{16,2}$, the first value indicates the cAIC value obtained on the subsample while the second one indicates the value obtained from refitting the model on the full sample.} 
\label{Tab:Pred1}
\end{table}
 
\begin{table}[H]
\centering
\tiny 
\begin{tabularx}{\textwidth}{p{0.7cm}P{3.2cm}P{3.1cm}P{3cm}P{3cm}P{2cm}P{1cm}}
\toprule
 & $\bm{B_{j,l}}$ & $\bm{C_{j,l}}$ & $\bm{D_{j,l}}$ & $\bm{E_{j,l}}$ & \textbf{p-value} & \textbf{cAIC} \\  
              \midrule
  \rowcolor{Gray}
  $A_{1,5}$ & M2 Gr. (\%), GDP Gr. (\%) & Fin. Open., Trade Open. (\% GDP), Year &  & & 1.00 &-861.16 \\ 
  $A_{2,5}$ & Trade Open. (\% GDP),  GDP Gr. (\%) & M2 Gr. (\%), Fin. Open., Year & Trade Open. (\% GDP) : Fin. Open. &  & 1.00  &-862.49 \\ 
    \rowcolor{Gray}
  $A_{3,5}$ & GDP Gr. (\%), & M2 Gr. (\%), Year & Fin. Open. : Trade Open. (\% GDP) & &  1.00 & -868.87\\ 
  $A_{4,5}$ &  Fin. Open., Trade Open. (\% GDP) & M2 Gr. (\%),  GDP pc (USD), Year &  & &  1.00 &-835.10 \\  
    \rowcolor{Gray}
  $A_{5,5}$ &  Fin. Open., Trade Open. (\% GDP), M2 Gr. (\%) & GDP pc (USD), Year & Trade Open. (\% GDP) : Fin. Open. &  & 0.80 &-835.97 \\ 
  $A_{6,5}$ & M2 Gr. (\%) & GDP pc (USD), Year &  Fin. Open. : Trade Open. (\% GDP) & & 1.00 &-841.52 \\ 
    \rowcolor{Gray}
  $A_{7,5}$ & GDP Gr. (\%), Fin. Open., Trade Open. (\% GDP) & Credit (\% GDP) Gr. (\%) & & & 0.01 &-1052.44 \\ 
  $A_{8,5}$ & GDP Gr. (\%), Fin. Open., Trade Open. (\% GDP) & Credit (\% GDP) Gr. (\%) & Fin. Open. : Trade Open. (\% GDP)&  & 0.05 &-1056.61 \\ 
    \rowcolor{Gray}
$A_{9,5}$ & GDP Gr. (\%),  Fin. Open., Trade Open. (\% GDP) & Credit (\% GDP) Gr. (\%) & & & 0.00 &-1052.44 \\ 
$A_{10,5}$ & GDP pc (USD), Fin. Open., Trade Open. (\% GDP) & Credit (\% GDP) Gr. (\%) & & & 0.00 &-1016.03 \\ 
    \rowcolor{Gray}
$A_{11,5}$ &  Trade Open. (\% GDP) & Credit (\% GDP) Gr. (\%), Fin. Open., GDP pc (USD) & Trade Open. (\% GDP) : Fin. Open. &  & 0.01 &-1017.63 \\ 
$A_{12,5}$ & GDP pc (USD), Fin. Open., Trade Open. (\% GDP) & Credit (\% GDP) Gr. (\%) &  & & 0.01 &-1016.03 \\
    \rowcolor{Gray}
$A_{13,5}$ & GDP Gr. (\%) & KOF Global., M2 Gr. (\%), Year &  & & 1.00 &-860.21 \\ 
$A_{14,5}$ & & GDP pc (USD), KOF Global., M2 Gr. (\%), Year &  & & 0.91 &-810.54 \\
    \rowcolor{Gray}
$A_{15,5}$ & GDP Gr. (\%) & KOF Global., Credit (\% GDP) Gr. (\%) &  &  & 0.03 & -1062.65 \\ 
$A_{16,5}$ & GDP pc (USD), KOF Global. & Credit (\% GDP) Gr. (\%) &  &  & 0.02 & -1026.54 \\
    \rowcolor{Gray}
$A_{17,5}$ & GDP Gr. (\%) & ICT Capital (\%), M2 Gr. (\%), Year &  &  & 0.59  &-794.95 \\ 
$A_{18,5}$ & M2 Gr. (\%), GDP pc (USD) & ICT Capital (\%), Year &  & & 0.63 &-819.88 \\
    \rowcolor{Gray}
$A_{19,5}$ & GDP pc (USD), ICT Capital (\%) & Credit (\% GDP) Gr. (\%) &  &  &  0.00 &-1041.86 \\ 
$A_{20,5}$ & GDP Gr. (\%), ICT Capital (\%) & Credit (\% GDP) Gr. (\%) &  & &  0.00 & \textbf{-1083.88} \\
          \midrule
\rowcolor{Gray}
$A_{1,6}$ & Age 65 (\%), GDP Gr. (\%) & M2 Gr. (\%), Year & & & 1.00 & -1127.62 \\  
$A_{2,6}$ & Age 65 (\%), GDP pc (USD) & M2 Gr. (\%), Year & & & 0.08 & -1105.93 \\  
    \rowcolor{Gray}
$A_{3,6}$ & Age 65 (\%), GDP Gr. (\%) & Credit (\% GDP) Gr. (\%), Year & &  & 1.00 &-1142.30 \\ 
$A_{4,6}$ & Age 65 (\%), Credit (\% GDP) Gr. (\%) & GDP pc (USD) & & & 0.04 & \textbf{-1308.49} \\
    \rowcolor{Gray}
$A_{5,6}$ & Age 75 (\%), GDP Gr. (\%) & M2 Gr. (\%), Year & & & 1.00 &-1116.97 \\ 
$A_{6,6}$ & M2 Gr. (\%), GDP pc (USD) & Age 75 (\%) & & & 0.00 &-1278.86 \\ 
    \rowcolor{Gray}
$A_{7,6}$ & Age 75 (\%), GDP Gr. (\%) & Credit (\% GDP) Gr. (\%), Year & & & 1.00 &-1132.36 \\ 
$A_{8,6}$ & Age 75 (\%), GDP pc (USD) & Credit (\% GDP) Gr. (\%) & & & 0.00 &-1264.68 \\ 
          \midrule
  \rowcolor{Gray}
$A_{1,7}$ & GDP Gr. (\%) & En. Prices (USD), En. Rents (\% GDP), M2 Gr. (\%), Year &  & & 1.00 & -1273.53 \\ 
$A_{2,7}$ & GDP Gr. (\%) & En. Prices (USD), En. Rents (\% GDP), M2 Gr. (\%), Year &  & En. Prices (USD) / En. Rents (\% GDP)  & 1.00 & -1304.82 \\ 
        \rowcolor{Gray}
$A_{3,7}$ & GDP Gr. (\%) & M2 Gr. (\%), Year &  & En. Prices (USD) / En. Rents (\% GDP) & 1.00 & -1266.08 \\ 
$A_{4,7}$ & En. Prices (USD), En. Rents (\% GDP), GDP pc (USD) & M2 Gr. (\%) &  &   & 0.00 & -1289.50 \\  
        \rowcolor{Gray}
$A_{5,7}$ & En. Prices (USD), En. Rents (\% GDP), GDP pc (USD) & M2 Gr. (\%) &  &  & 0.00 &  -1289.50 \\   
$A_{6,7}$ & En. Prices (USD), En. Rents (\% GDP), GDP pc (USD) & M2 Gr. (\%) &  &  & 0.00 & -1289.50 \\  
        \rowcolor{Gray}
$A_{7,7}$  & GDP Gr. (\%) & En. Prices (USD), En. Rents (\% GDP), Credit (\% GDP) Gr. (\%), Year &  & & 1.00 &  -1317.56 \\ 
$A_{8,7}$  & GDP Gr. (\%) & En. Prices (USD), En. Rents (\% GDP), Credit (\% GDP) Gr. (\%), Year &  & En. Prices (USD) / En. Rents (\% GDP)  &1.00 &  -1321.54 \\ 
        \rowcolor{Gray}
$A_{9,7}$  & GDP Gr. (\%) & Credit (\% GDP) Gr. (\%), Year &  & En. Prices (USD) / En. Rents (\% GDP)  & 1.00 & -1351.89 \\ 
$A_{10,7}$ & En. Prices (USD), En. Rents (\% GDP), GDP pc (USD) & Credit (\% GDP) Gr. (\%) &  & &   0.00 & -1279.86 \\ 
        \rowcolor{Gray}
$A_{11,7}$ & En. Rents (\% GDP), Credit (\% GDP) Gr. (\%), GDP pc (USD) & En. Prices (USD) & &  & 0.00 &-1468.50 \\ 
$A_{12,7}$ & & Credit (\% GDP) Gr. (\%), GDP pc (USD) & & En. Prices (USD) / En. Rents (\% GDP) & 0.00 & \textbf{-1544.34} \\ 
        \midrule
  \rowcolor{Gray}
$A_{1,8}$ & GDP Gr. (\%) & M2 Gr. (\%), Past Infl. (\%) & & & 0.00  &-1283.38 \\ 
$A_{2,8}$ & Past Infl. (\%), GDP pc (USD) & M2 Gr. (\%) & & & 0.00 &\textbf{-1294.65}\\ 
\rowcolor{Gray}
$A_{3,8}$ & GDP Gr. (\%) & Credit (\% GDP) Gr. (\%), Past Infl. (\%), Year & & & 1.00 &-1150.11 \\ 
$A_{4,8}$ & GDP pc (USD), Past Infl. (\%) & Credit (\% GDP) Gr. (\%), Year & & & 1.00 &-1132.82 \\ 
    \midrule
  \rowcolor{Gray}
  $A_{B}$ & Debt (\% GDP) Gr. (\%), Debt (\% GDP), Trade Open. (\% GDP), M2 Gr. (\%) & Past Infl. (\%), Year, En. Prices (USD), En. Price Gr. (\%), Credit (\% GDP), Out. Gap (\%), Fin. Open.  & & En. Prices (USD) / En. Rents (\% GDP) & 0.00 & \textbf{-1700.26} \\ 
\bottomrule
\end{tabularx}
\caption{(2/2) Allocation of the predictor set $A_{j,l}$ of the model $M_{j,l}$ to $B_{j,l}$,$C_{j,l}$,
$D_{j,l}$ and $E_{j,l}$. The cAIC value for the AMM with the lowest cAIC is printed in bold. The model complexity of $M_{9,5}$, $M_{12,5}$, $M_{5,7}$ and $M_{6,7}$ had to be reduced (i.e. fewer model parameters) due to non-convergence during the optimization of the initially intended AMM.} 
\label{Tab:Pred2}
\end{table}

\clearpage

\begin{figure}[H]
    \centering
    \includegraphics[width = \textwidth]{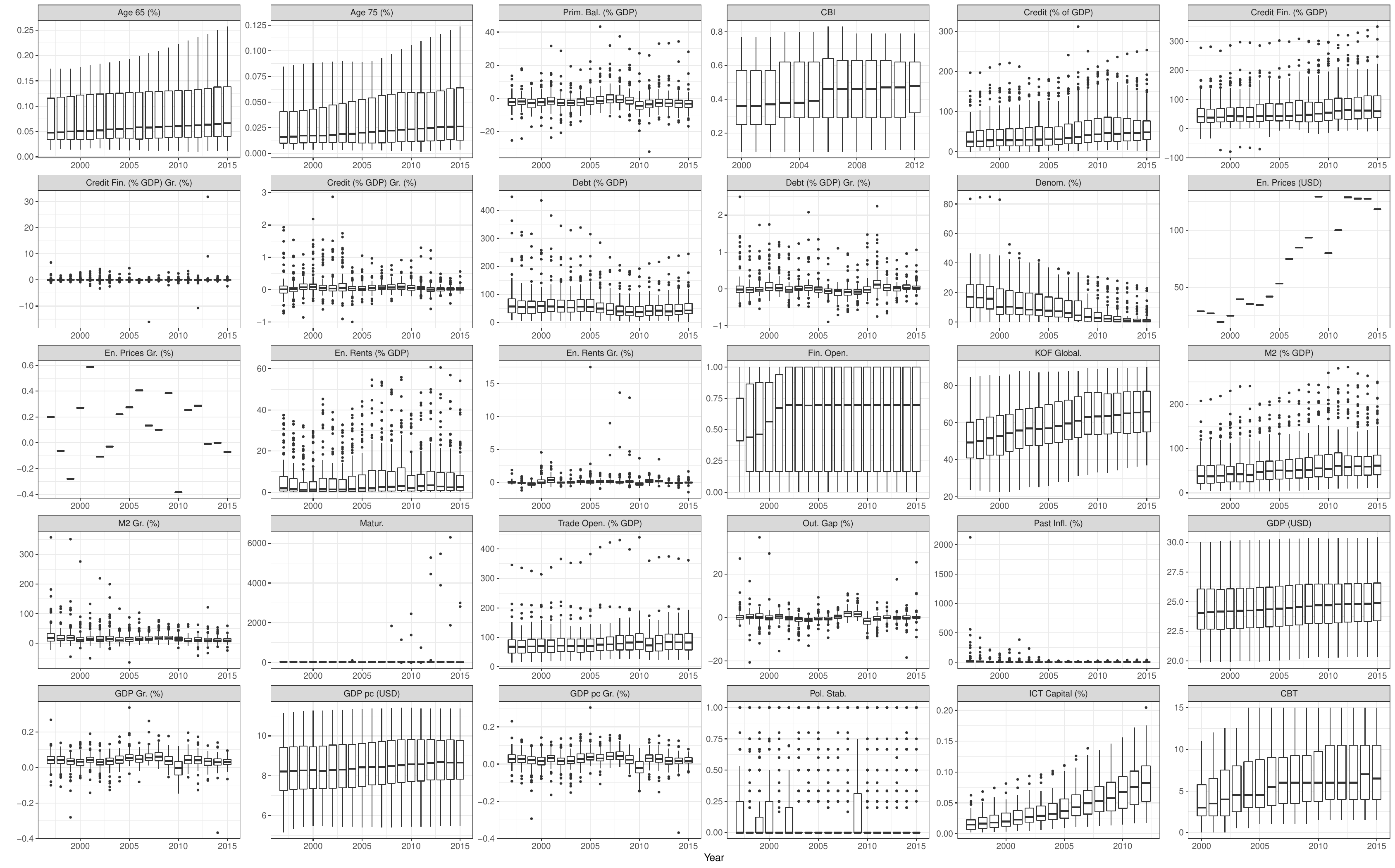}
    \caption{Descriptive statistics for all metric predictors comprised by the data set. The variables GDP (USD) and GDP pc (USD) have been transformed by the natural logarithm for visualization.}
    \label{fig:Numerics}
\end{figure}

\begin{figure}[H]
    \centering
    \includegraphics[width = \textwidth]{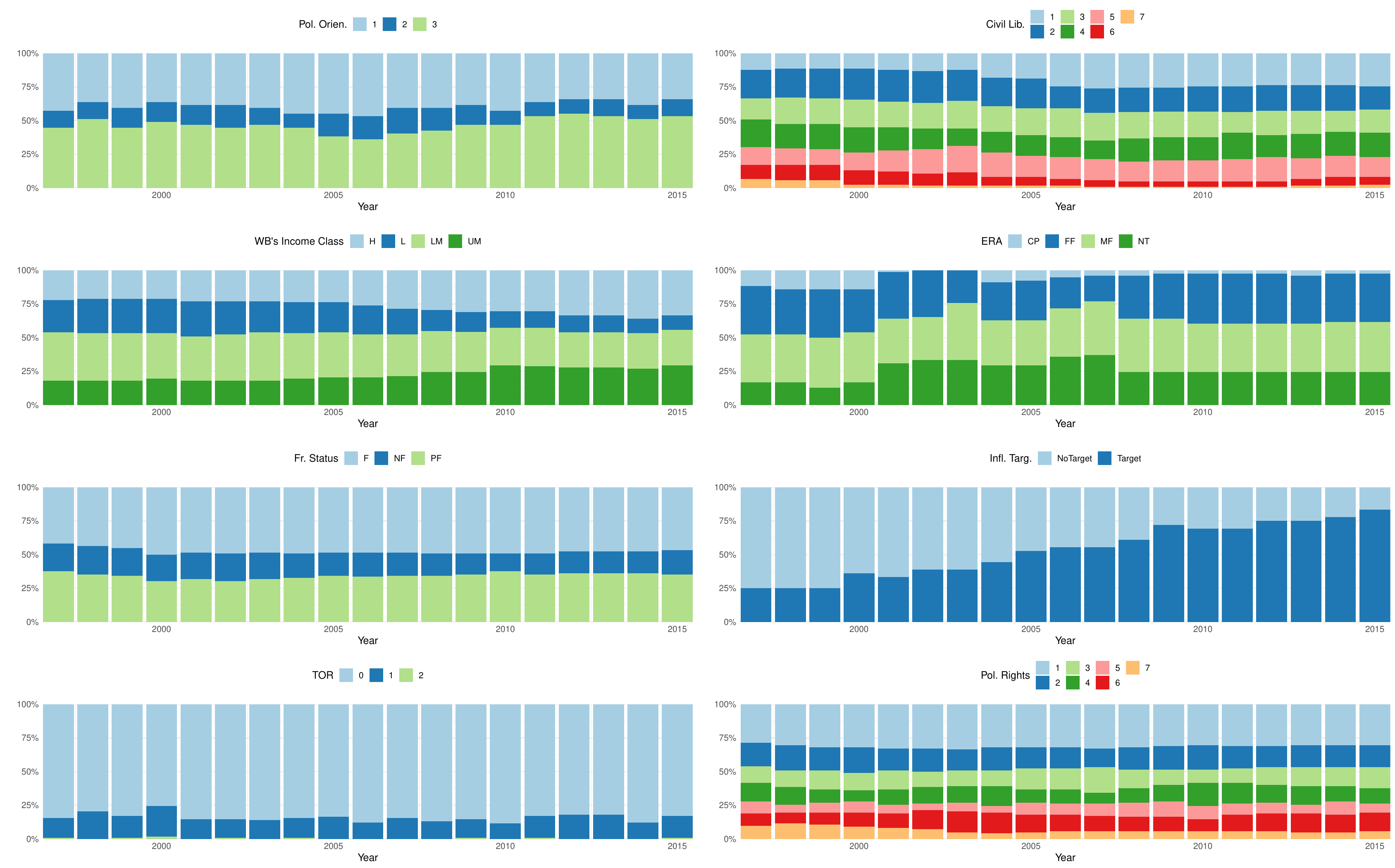}
    \caption{Descriptive statistics for all categorical predictors comprised by the data set and the World Bank's income classification.}
    \label{fig:Cats}
\end{figure}

\section{Appendix: Methods}\label{sec:AppendixMethods}

\subsection{Mundlak Test}\label{sec:MundlakAppendix}

We obtained the conditional (cond.) log-likelihood as specified in Section \ref{sec:cAIC} based on \eqref{eq:MundlakAMM}, $l_A$, and based on \eqref{eq:AMM}, $l_0$, which then enabled us to compute the test statistics
\begin{equation}
    T_{LRT} = 2 \sup_{H_A} l_A - 2 \sup_{H_0} l_0
\end{equation}
This test statistics is asymptotically $\chi^{2}_{rank(L)}$ distributed where $L$ is the matrix of contrasts that varies with each $M_{j,l}$.

\subsection{cAIC}\label{sec:cAIC}
Model selection based on the Akaike information criterion (AIC) is a common approach in econometrics. The criterion was initially introduced by \citet{akaike1973information} and is composed of twice the maximized log-likelihood and a bias correction term, which, under certain regularity conditions, can be estimated asymptotically by two times the dimension of the unknown parameter vector specifying this log-likelihood \citep{safken2018conditional}. To apply this criterion to mixed models, two considerations were taken into account in our case.

First, a joint Gaussian distribution of the random vectors $\bm{y}$ and $\bm{b}$ is assumed. This allows us to decide between two common views regarding the inference and predictions in mixed models. The distribution of $\bm{y}$ cond.\ on $\bm{b}$ leads to the cond.\ likelihood of $\bm{y}$ given $\bm{b}$, which then forms one component of our utilized cAIC. In contrast, when the random effects are integrated out, the marginal distribution of $\bm{y}$ emerges and thus provides the marginal likelihood. We demonstrate our reasoning for the conditional over the marginal view on the AIC in Section \ref{sec:ConditionalViewcAIC} in the Appendix.

Second, the bias correction term needs to be adjusted owing to the alternated number of parameters estimated in mixed models. A body of literature \citep{vaida2005conditional, liang2008note, greven2010behaviour} provides the theoretical underlying for the derivation of the bias correction. We give a brief overview in Section \ref{sec:BackgroundcAIC} in the Appendix. We base our analysis on the term introduced by \citet{greven2010behaviour} (cf. \eqref{eq:cAIC} in the Appendix), which assumes independent and identically distributed errors across the subjects (countries in our case). As a result, its current software implementation in the \texttt{cAIC4} package \citep{cAIC4manual} originally provided for mixed models emerging from the \texttt{lme4} package \citep{JSSv067i01} and the \texttt{gamm4} package \citep{wood2016generalized} incorporates this assumption as well. However, in our case subject-specific error variances need to be modeled to capture the heterogeneity across countries, making the assumption of identically distributed errors inappropriate. The derived bias correction is thus no longer applicable since it disregards the additional parameters used for the estimation of $\bm{R}$ as defined in Section \ref{sec:MethodsAMM}. In order to account for the estimation of a more complex error covariance structure in the bias correction, we incorporate the proposed extension of \citet{overholser2014effective}. Since \citeauthor{overholser2014effective} do not take into account the estimation uncertainty of $\bm{G}$, we implemented a working version that adds the number of unknown parameters $r$, which we used for the estimation of the error covariance matrix $\bm{R}$, to the bias correction term of \eqref{eq:cAIC} and obtained
\begin{equation}\label{eq:cAICwithR}
    \text{cAIC} = -2 \text{log} f(\bm{y}|\bm{\theta},\bm{b}) + 2 \left(\text{tr}\left(\frac{\partial \bm{\hat{y}}}{\partial \bm{y}}\right) + r\right).
\end{equation}
We implemented \citeauthor{overholser2014effective}'s proposal for diagonal error covariance matrices into the \texttt{cAIC4} package and further extended the package for mixed and additive models estimated with the \texttt{mgcv} package \citep{mgcv2011}. As a result, we provide, to our knowledge, the first software implementation for the estimation of the cAIC for mixed and additive models with non identically distributed errors. This novel extension of \texttt{cAIC4} is made available to the CRAN repository for further applications. The proof of the asymptotic result of \citet{overholser2014effective} gives an upper bound for the bias correction term that can also be provided through derivations based on the partial derivative of the prediction vector $\bm{\hat{y}}$ for $\bm{\bm{y}}$ with the random effects set to their predicted values.

\subsection{Conditional Over the Marginal View on the AIC}\label{sec:ConditionalViewcAIC}
We prefer the conditional over the marginal perspective on the AIC due to the mixed model representation of P-splines. To see the link in general, following \citet{safken2018conditional}, we consider an additive model of the following form 
\begin{equation}
\bm{y} = \bm{Ba} + \bm{\varepsilon}, \quad \bm{\varepsilon} \sim N(\bm{0},\sigma^2 \bm{I})
\end{equation}
where $\bm{B}$ is the design matrix containing the evaluations of predictors based on B-spline basis functions constructed from piece-wise polynomials and $\bm{a}$ is the corresponding vector containing the basis coefficients. We can apply an eigenvalue decomposition to the quadratic penalty matrix $\bm{P} = \bm{D}^\top \bm{D}$ with column rank $k$, where $\bm{D}$ is the differences matrix such that $\bm{P}=\bm{V} \Tilde{\bm{\Sigma}} \bm{V}^\top$.  The $k$ eigenvectors in $\bm{V}$, which correspond to the $k$ positive eigenvalues, can be assigned to $\bm{V_1}$ while the remaining $d$ column vectors can be assigned to $\bm{V_0}$. We are then able to non-uniquely decompose the functional estimate $\bm{Ba}$ into two bases
\begin{equation}\label{eq:MMAdditive}
\bm{Ba} = \bm{BV_0V_0^\top a} + \bm{BV_1V_1^\top a} = \bm{X\beta} + \bm{Zb}
\end{equation}
yielding the common mixed model representation \citep{verbyla.1999}, where $\bm{\beta}$ specifies $d$ unpenalized parameters with the corresponding fixed effects design matrix $\bm{X}$ spanning the polynomial null space of $\bm{P}$, while $\bm{b}$ specifies $k$ penalized parameters corresponding to the random effects design matrix $\bm{Z}$ which spans its complement \citep{Eilers.2015}, respectively. \citet{bivariateMMSplines} extend this representation for penalized functionals in higher dimensions. The specification under \eqref{eq:MMAdditive} differs from our generic mixed model \eqref{eq:AMM} by different column ranks of the fixed and random design matrices and different dimensions of the corresponding parameter vectors, depending on the employed predictor sets $A_{j,l}$ as defined in Section \ref{sec:MethodsAMM}. As a result, with $d = 2$ in our case, the marginal AIC would only take the fixed polynomial trend of degree one into account while the  smooth deviation from this polynomial can now be taken into account in \eqref{eq:cAICwithR} as well. Thus, the cAIC, considered as a predictive measure in this context, accounts for the plausible assumption that the non-linear functional relation between the predictors and $\bm{y}$ estimated in our data set represents a more general relationship which is expected to hold also for new country observations.\footnote{See \citep{greven2010behaviour,safken2018conditional}.}

\subsection{Background on the cAIC}\label{sec:BackgroundcAIC}

Following the introduction of the cAIC by \citet{vaida2005conditional} the bias correction term is defined as
\begin{equation}\label{eq:BiasCorrection}
    \text{BC} = 2 tr(\bm{H})
\end{equation}
where $\bm{H}$ is the hat matrix projecting $\bm{y}$ onto $\bm{\hat{y}}$ where $\bm{\hat{y}}$ is a prediction vector for $\bm{y}$ with the random effects set to their predicted values. However, their proposal assumes known variance parameters and neglects the estimation uncertainty of these variance parameters. This estimation uncertainty can be taken into account whereby the bias correction term depends on the assumed cond.\ distribution of $\bm{y}$. For our Gaussian case, \citet{liang2008note} define
\begin{equation}\label{eq:cAIC}
    \text{cAIC} = -2 \text{log} f(\bm{y}|\bm{\theta},\bm{b}) + 2 \text{tr}\left(\frac{\partial \bm{\hat{y}}}{\partial \bm{y}}\right)
\end{equation}
where $\bm{\theta}$ is defined as the vector of parameters in the model. \citet{liang2008note} approximate the trace of the derivatives of the estimated and predicted quantities numerically. However, this becomes computationally infeasible at a moderate sample size and a large quantity of models similar to in our case \citep{greven2010behaviour}. \citet{greven2010behaviour} provide closed form expressions for these derivatives, circumventing the numerical approximations yielding an analytic representation of the cAIC which takes estimation uncertainty of the variance parameters into account. \citet{safken2018conditional} offer an efficient software implementation by means of the add-on package \texttt{cAIC4} \citep{cAIC4manual}.

\subsection{Model-based Boosting}\label{sec:BoostingAppendix}

\paragraph{Loss Function} The boosting algorithm minimizes the empirical risk which is given by
\begin{equation}\label{eq:boostingRisk}
\Ra := \frac{1}{n}\sum_{i=1}^n \La(y_i, f(\boldsymbol{x_i}))
\end{equation}
where ($y_i$,$\boldsymbol{x_i}$) is one out of $i = 1,\ldots,n$ realizations of ($\bm{y}$,$\boldsymbol{x}$). The Huber-Loss, $\La$, was chosen because of its advantages in handling outliers compared with other approaches. The Huber-Loss is defined as
\begin{align}
    \La(\bm{y}, f; \delta) = \begin{cases}
    \frac{1}{2}(\bm{y}-f)^2 & \text{for} \ |\bm{y}-f| \leq \delta, \\
    \delta(|\bm{y}-f|-\frac{\delta}{2}) & \text{for} \ |\bm{y}-f| > \delta
    \end{cases}
\end{align}
and $\delta$ was chosen in each boosting iteration $m$ by
\begin{align*}
	\delta_m = \text{median}(|y_i - \hat{f}_{m-1}|, i = 1,...,n)
\end{align*}

\paragraph{Base Learners} All predictors specified in Section \ref{sec:lit} which are available in the full sample were collected in $\bm{x} := (\bm{x}_1, \ldots, \bm{x}_p)$. For the case of inflation, four kinds of base learners are specified. The first type are penalized least squares base learners which model all categorical predictors in $\bm{x}$. The second type are P-spline base learners which model all continuous predictors in $\bm{x}$. The third type are bivariate P-splines base learners allowing for the estimation of smooth interaction surfaces. We allow for the same bivariate interactions of predictors as we have done for the models specified by economic theory -- \textit{En. Prices (USD)} and \textit{En. Rents (\% GDP)} denoted by $f_{1,2}$ and \textit{Trade Open. (\% GDP)} and \textit{Fin. Open.} denoted by $f_{3,4}$. The last type are random effect base learners for country-specific random intercepts, $f_{intercept}$, and country-specific random slopes, $f_{slope}$, with Ridge-penalized effects. We finally add a global intercept such that the following additive model results
\begin{equation}\label{eq:boostingFormula}
\mathbb{E}[\bm{y}|\bm{x}] = \beta_0 + f_1 + \ldots + f_p + f_{1,2} + f_{3,4} + f_{intercept} + f_{slope}
\end{equation}

\paragraph{Gradient Descent Algorithm} The utilized gradient boosting algorithm starts with an initial function estimate $\hat{f}_{[0]}$ and proceeds in a stagewise manner. At each iteration $m$ it computes the negative gradient of the loss function and updates the current function estimate $\hat{f}^*_{[m]}$. Simultaneously, the algorithm descends along the gradient of the empirical risk $\Ra$ whereby only one base learner is selected at each iteration for updating the current function estimate. The decision when to stop the algorithm, $m_{stop}$, is crucial. 
However, it has been commonly suggested to enforce a stop of the algorithm before it converges to avoid overfitting and thus a suboptimal prediction \citep{Buehlmann.2007}.

\paragraph{Selection Procedure} We employ a 10-fold bootstrap to find $m_{stop}$ by choosing the minimum out-of-sample risks averaged over all folds \citep{mboost}. To take the longitudinal structure of our data into account this procedure was stratified by countries. 
To enforce variable selection, we decided to include only the base learners that were selected at least 1\% of all $m_{stop}$ iterations.

\subsection{Varying Coefficient Models}\label{sec:VarCoef}

After the model selection procedure, we additionally answered the important question of a structural break for the parameters comprised by $M^{*}_{l}$, $l = 1,\ldots,8$ and $M_{B}$ through varying coefficient models. That is, we let each parameter interact with a two-level categorical variable $e_{i,t}$ such that \eqref{eq:LinPred} was replaced by
\begin{align}
\Tilde{\eta}_{i,t} = \sum_{x \in  B_{j,l}} x_{i,t} \beta_{x} e_{i,t}
+ \sum_{(x,x^{*}) \in D_{j,l}}  (x_{i,t}x_{i,t}^{*}) \beta_{(x,x^{*})} e_{i,t} \nonumber \\
+ \sum_{x \in C_{j,l}} h_x(x_{i,t}) e_{i,t}
+ \sum_{(x,x^{*}) \in E_{j,l}}  f_{(x,x^{*})}(x_{i,t},x_{i,t}^{*})  e_{i,t} \label{eq:VaryCoef}
\end{align}
for each $M^{*}_{l}$, $l = 1,\ldots,8$ and $M_{B}$. The first level of $e_{i,t}$ is considered when $ t \leq 2007 $ and the second level when $ t > 2007 $.  Consequently, we obtain two simultaneous estimations of the same effect -- one for each of the two levels. However, apart from the specification of \eqref{eq:VaryCoef}, every model specification was identical to the model specification of the original $M^{*}_{l}$, $l = 1,\ldots,8$ or $M_{B}$ respectively.

\subsection{Motivation for AMMs}\label{sec:MotivationAMM}

To motivate the use of AMMs we now demonstrate the pitfalls that come along with polynomial regressions of order two which are frequently used to study potential non-linear effects on inflation. However, these pitfalls also arise from polynomial regressions of higher order. To show this, we replicate Table 1 of \citet{HielscherMarkwardt2012} and highlight the limitations that go hand in hand with the parametric assumptions underlying their results. We chose the study by \citet{HielscherMarkwardt2012} for three reasons. First, this study makes an important contribution to the literature and is often cited. Second, it provides one of the few analyses of potential non-linear effects of inflation. Third, it has important policy implications suggesting that an increase in CBI does not reduce inflation unless the change in CBI is sufficiently large, and the quality of political institutions is sufficiently high. To replicate \citeauthor{HielscherMarkwardt2012}s' paper we collected inflation rates\footnote{These rates are identical to the inflation rates used in our study.} for the 69 countries analyzed by the authors from the IMF and complemented them with the World Bank's inflation rates for Argentina and Nigeria which are missing in the IMF data base. For Serbia's inflation rate between 1980 and 1991 we relied on \href{https://www.cato.org/commentary/worlds-greatest-unreported-hyperinflation}{CATOs} estimate for Yugoslavia's annualized inflation rate estimated at 76\%. For the CBI measure we recurred to the legal index constructed by \citet{Garriga2016} which is based on the index of \citet{Cukierman1992} employed by \citet{HielscherMarkwardt2012}. We aggregated these longitudinal data along the time dimension as reported by \citeauthor{HielscherMarkwardt2012} employing the geometric mean in the case of inflation and the arithmetic mean in the case of the CBI index. The results are in Table \ref{tab:RepStudy}. We were able to replicate the sign of the coefficients in all cases, whereas the significance level could be replicated in 14 out of 22 cases.

\begin{table}[ht]
\centering
\begin{tabularx}{\textwidth}{YYYYYYY}
  \hline
& (1) & (2) & (3) & (4) & (5) & (6) \\ 
  \hline
  \hline
Intercept & -0.064*** & -0.036*** & -0.04*** & -0.02** & -0.037*** & -0.017 \\ 
& (0.016) & (0.008) & (0.013) & (0.008) & (0.013) & (0.011) \\ 
  \hline
 $\pi_{\text{80/89}}$  & 0.977*** &  - & 0.985***  & - & 0.953*** & - \\
& (0.011) &  &  (0.012) &  &  (0.045) & \\
   \hline
 $\pi^{T}_{\text{80/89}}$  & - & 0.834*** & - & 0.86*** & - & 0.804*** \\ 
& & (0.062) & & (0.057) &  & (0.101) \\ 
  \hline
$\Delta$ CBI  &  -0.006 & -0.005 & -0.34** & -0.256** & -0.316** & -0.241** \\ 
&  (0.037) & (0.025) & (0.155) & (0.111) & (0.154) & (0.108) \\ 
  \hline
$\Delta$ $\text{CBI}^2$ & - & - & 0.551** & 0.415** & 0.526** & 0.401** \\ 
&   &  & (0.249) & (0.181) & (0.245) & (0.177) \\ 
  \hline
No. observ. &  69 & 69  & 69 & 69 & 66 & 66 \\ 
  \hline
Adj. $\text{R}^2$ &  0.97 & 0.73  & 0.97 & 0.74 & 0.89 & 0.6 \\ 
   \hline
\end{tabularx}
\caption{Replication of Table 1 of \citet{HielscherMarkwardt2012}. All columns with odd numbers refer to the change in inflation and all columns with even numbers refer to the change in inflation tax. In columns (5) and (6) all countries which experienced an absolute change in inflation of more than 200 percentage points are excluded. All standard errors are reported in parentheses. Estimates are based on White heteroscedasticity-consistent standard errors. Significance levels are reported as follows: * for a 10\%, ** for a 5\% and *** for a 1\%-significance-level.}
\label{tab:RepStudy}
\end{table}

We next show how the estimation of \citet{HielscherMarkwardt2012} can be improved by AMMs as well as the additional insights that can be gained from this procedure. For this purpose, we focus on the inflation tax variable $\frac{\pi}{1 + \pi}$ (columns 2, 4 and 6 of Table \ref{tab:RepStudy}) which is more robust to outliers. Column 2 provides evidence of a positive linear effect of the initial inflation tax $\pi^{T}_{\text{80/89}}$. However, modeling $\pi^{T}_{\text{80/89}}$ by means of P-splines based on the model specification illustrated in column 2, whose results are exhibited in Figure \ref{fig:p_spline_inflation_tax} below, shows compelling evidence of a strong non-linear effect on the change in inflation tax (EDF of $15^{***}$) and an increase in the adj. $R^2$ from 0.73 (column 2) to 0.86\footnote{Results are available upon request.}. Figure \ref{fig:p_spline_inflation_tax} exhibits not only a non-linear effect but even suggests a negative effect from the initial inflation tax on the change in the inflation tax between 0.2 and 0.3. This contrasts with the positive effects reported in Table \ref{tab:RepStudy}. Modeling  $\pi^{T}_{\text{80/89}}$ by a linear and a quadratic term increases the adj. $R^2$ by only 0.02. This means that linear and quadratic terms are too restrictive parametric assumptions that limit the explanatory power of the initial inflation tax variable. Further, modeling $\pi^{T}_{\text{80/89}}$ by means of P-splines based on the specification exhibited in column 4 shows the same evidence as the regression based on column 2 (EDF of $15.3^{***}$ and adj. $R^2$ of 0.86). Furthermore, by taking the non-linearity of $\pi^{T}_{\text{80/89}}$ into account, the effect of $\Delta$ CBI on the inflation tax is even reduced ($\Delta$ CBI: -0.186; $\Delta$ $\text{CBI}^2$: 0.265). This pattern persists even after excluding the high-inflation countries as in column 6. Here, the EDF increases to a value of $16^{***}$ and the adj. $R^2$ increases from 0.60 to 0.83, again mitigating the effect of $\Delta$ CBI ($\Delta$ CBI: -0.182; $\Delta$ $\text{CBI}^2$: 0.261). 

Summarizing our evidence, we emphasize three points. First, polynomial regressions, which come mostly in the form of quadratic modeling, may result in too restrictive parametric assumptions and reduce the explanatory power of the associated economic theory. Second, this type of model misspecification may suggest fallacious policy implications. Third, misspecification of single regressors have effects on the remaining regressors, exacerbating and spreading the problem of incorrect conclusions. As we were able to show, all these critical issues can be addressed by AMM. 

\begin{figure}
    \centering
    \includegraphics[width=\textwidth]{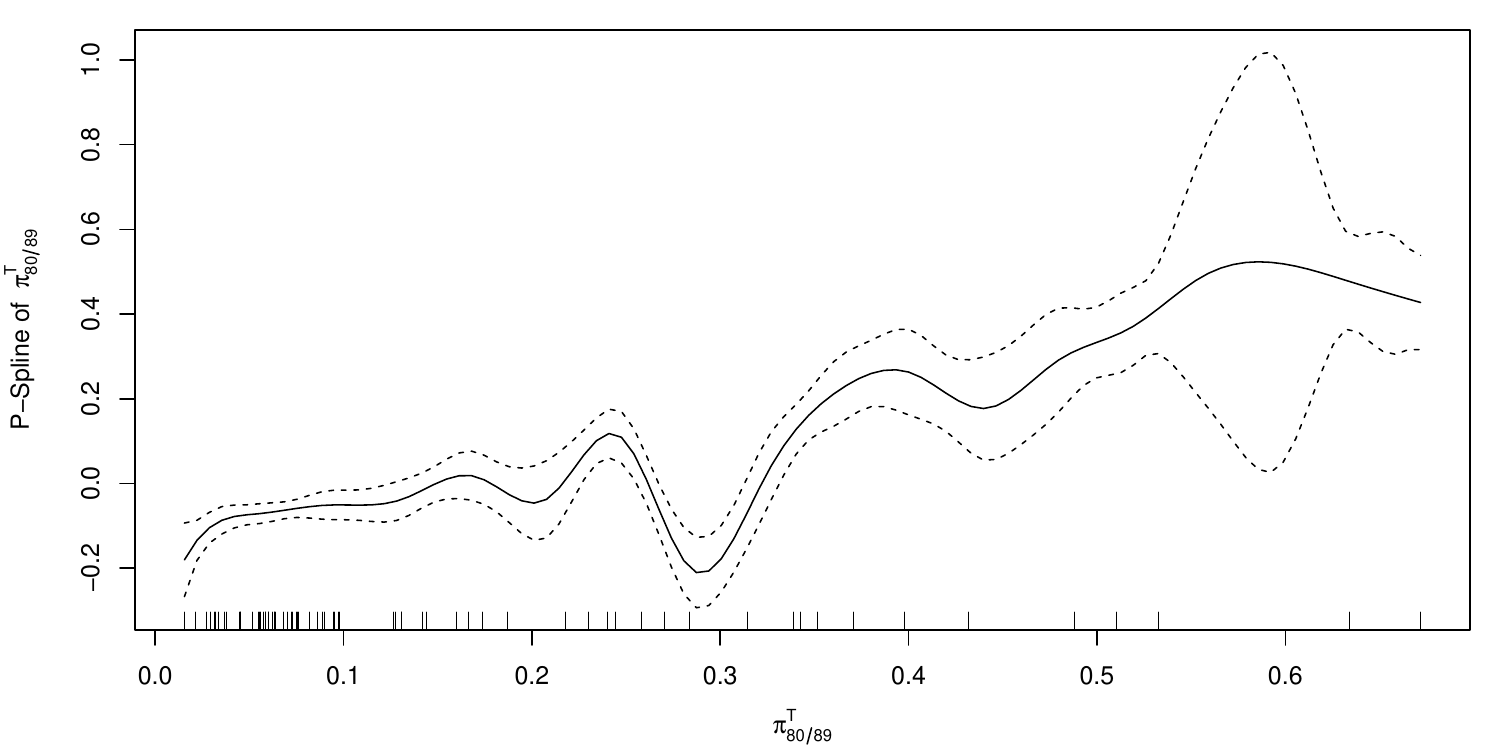}
    \caption{Estimated effect of the non-linear modeling of the $\pi^{T}_{\text{80/89}}$ variable. Every other specification is identical to column (2) of Table \ref{tab:RepStudy}.}
    \label{fig:p_spline_inflation_tax}
\end{figure}

\section{Appendix: Literature}\label{sec:ResultsLiterature}

\subsection{Money, Credit, and Slack}

A key macroeconomic axiom is the quantity theory of money. It posits a proportional relation between the growth rate of money and inflation. Numerous studies confirm that sustained high growth rates of money in excess of its production of goods and services eventually produce high and rising inflation rates. The quantity theory does not specify which definition of the money supply should be used in empirical tests. We measure money supply by the growth rate of M2 (\textit{M2 Gr. (\%)}).

In addition to money we also examine the effect of credit creation. Two opposite effects are possible -- an inflationary and a disinflationary one. On the one hand, an inflation-raising effect may arise from credit expansions that go hand in hand with money creation. On the other, domestic credit may proxy financial depth and, as such, contain information expected to be negatively related to inflation.\footnote{See \cite{CalderonSchmidtHebbel2010}.} In addition, a credit expansion that leads to a build-up of investment and an expansion of production capacities put downward pressure on prices. We employ two measures of domestic credit. The first refers to credit to the private sector in percent of GDP (\textit{Credit (\% GDP)}) and the second to its growth rate (\textit{Credit (\% GDP) Gr. (\%)}). 

Another central tenet of macroeconomics is that the real and nominal sides of the economy are linked through a Phillips curve relationship. In the New Classical form, inflation is a function of lagged expected inflation and a contemporaneous measure of excess demand or slack, often measured by the output gap, defined as actual minus potential output. In the New Keynesian model, current inflation is related to (rationally) expected future inflation, along with a measure of demand. An important question is whether the financial crisis of 2008-2009 led to structural breaks. One particular point in case concerns the Phillips curve. The background is the missing disinflation which was observed in advanced economies in the aftermath of the financial crisis.\footnote{For recent reviews of papers on the apparent flattening of the U.S. Phillips curve in the 2000s, and especially since the financial crisis, see \cite{McLeayTenreyro2019} and \cite{HooperMishkinSufi2019}.} We calculate potential or trend output by a Hodrick-Prescott (HP) filter-based measure with a lambda set to 6.25. The resulting variable is denoted as \textit{Out. Gap (\%)}.

\subsection{Institutions}

The potentially numerous motives for authorities to inflate may be prevented by sound institutions. We split a country's institutional setup in five separate items---central bank independence, central bank transparency, political instability and orientation, civil rights, and economic growth. 

Central bank independence has received much attention in explaining cross-country differences in inflation rates. We use the Dincer-Eichengreen index of de jure central bank independence (\textit{CBI}\@), which ranges from 0 (most dependent) to 1 (most independent) available from 1998 to~2010. In addition, we measure the occurrence of turnovers at the headquarters of the central bank during the year (\textit{TOR}) \citep{TOR}. 

One further fact that needs to be accounted for in the analysis of central bank independence is that its increase has been accompanied by greater transparency. Overall, the empirical literature documents beneficial effects from transparency. For example, \cite{DincerEichengreen2014} found that greater transparency is associated with lower average levels of inflation. We measure central bank transparency (\textit{CBT}) by the updated values of Dincer and Eichengreen that extend the observations reported in \cite{DincerEichengreen2014} by four more years until 2014.\footnote{We downloaded the updated version of February 2017.}

Political instability adds another potentially important element to the analysis of institutions for inflation. Various measures for political (in)stability have been applied. We rely on two. The first, \textit{Pol. Stab.}, measures the percentage of cabinet members (``veto players”) dropping out in any given year. The variable ranges from 0 (most stable) to ~1 (most unstable). The second is the party orientation with regard to their economic policy, \textit{Pol. Orien.}, categorized into three classes: left, center, and right. Both variables are obtained from the World Bank's database of political institutions. 

Another arguably important institutional element is political rights which we approximate with three categorical variables taken from \citet{freedomhouse.2015}: political rights (\textit{Pol. Rights}) taking on seven levels, civil liberties (\textit{Civil Lib.}) also seven levels, and freedom status (\textit{Fr. Status}) with three levels.

Finally, GDP and its growth rate, respectively, are considered important inflation-related factors that many authors have used. GDP per capita has been used as a proxy for institutional quality. Consequently, we rely on two forms, real GDP per capita (\textit{GDP pc (USD)}) and real GDP growth (\textit{GDP Gr. (\%)}).

\subsection{Monetary Policy Strategies}

The next group of variables accounts for the effects on inflation associated with two monetary policy arrangements. The first relates to exchange rate arrangements, the second to adopting an explicit inflation targeting (IT) strategy. One rationale for adopting a fixed exchange rate framework is that it operates as a disciplinary tool for monetary authorities. Another benefit is that a fixed exchange rate signals enhanced credibility of lower future inflation. As a result, inflation should be lower in countries with fixed exchange rates. Exchange rate arrangements come in a variety of forms. For our analysis, we extend the Reinhart-Rogoff classification. The resulting variable (\textit{ERA}) distinguishes four categories: no separate legal tender, a crawling peg, managed floating, and free-floating. Inflation targeting is an operational framework for monetary policy aimed at achieving a numerical value (or range) for the inflation rate. A growing number of countries have adopted IT over the last two decades. We created a binary variable, \textit{Infl. Targ.}, that takes the value of ~1 for countries that have adopted~IT, and ~0 otherwise. For its construction, we relied on various annual reports on Exchange Arrangements and Exchange Restrictions provided by the IMF.

\subsection{Public Finances}

A well-established theory in macroeconomics is that governments running persistent deficits have sooner or later to finance those deficits with seigniorage. In the aftermath of the financial crisis, financial bailouts, stimulus spending, and lower tax revenues have resulted in public debts in advanced economies that have surpassed the peaks reached during World War I and the Great Depression \citep{ReinhartRogoff2011}. There are several theoretical channels for how public indebtedness may unleash inflation: Conventional view, Unpleasant Monetarist Arithmetic (UMA), Fiscal Theory of the Price Level (FTPL), Optimal Tax and Debt Management. 

According to the conventional view, an increase in public debt may cause inflation by inducing a positive wealth effect on households \citep{ElmendorfMankiw1999}. The main result of the seminal paper on UMA by \cite{SargentWallace1981} is that the effectiveness of monetary policy in controlling inflation depends critically on its coordination with fiscal policy. A similar reasoning lies behind the  FTPL. Government debt not backed by expected future surpluses ensues in inflation, immediately or---depending on the maturity structure---in the future \citep{Cochrane2001}. A fourth explanation is based on the Theory of Optimal Taxation, which holds that governments optimally equate the marginal cost of the inflation tax with that of output taxes.\footnote{See \cite{Phelps1973}, \cite{Vegh1989}, and \cite{Aizenman1992}.} A potentially important issue is also the public debt structure. In the models by \cite{Calvo1988} and \cite{MissaleBlanchard1994}, higher levels of privately held government debt with a longer nominal maturity raise the incentive for a government to attempt surprise inflation. In this literature, foreign currency, inflation-indexed, or short-term debt are remedies against surprise inflation. Following the literature, we use different measures of fiscal stance. One captures the primary balance, \textit{Prim. Bal. (\% GDP)}. Another relates to debt growth (\textit{Debt (\% GDP) Gr. (\%)}). As proxies for testing the implications of theories on public debt management, we use average maturity on new external debt (\textit{Matur.}) as well as the percentage of external long-term public and publicly guaranteed (PPG) debt contracted in multiple currencies for the low- and middle-income countries (\textit{Denom. (\%)}).

\subsection{Globalization and Technology}

Declining inflation in many countries over the past few decades simultaneously as rising global competition has led to a debate on the importance of globalization for domestic inflation. We proxy openness by three variables. One refers to economic openness (\textit{Trade Open. (\% GDP)}), and reflects the sum of exports and imports divided by GDP (the most commonly used measure of openness). The second measures the openness in the capital account (\textit{Fin. Open.}; \citet{chin.2006}). The third openness proxy is the KOF Globalization index \citep{gygli2019kof} which we denote as \textit{KOF Global.} In addition to globalization, a view embraced by several authors is that technological progress may lead to declining prices of information and communications technology (ICT) products by reducing entry barriers for new producers, and by lowering wage growth. We examine the impact of technology by the ICT capital proxy of \cite{JorgensonVu2005} used by \cite{Jaumotteetal2013} and denote it as \textit{ICT Capital (\%)}.

\subsection{Demography}

In an effort to understand the sources of the decline in inflation observed over the recent past, the adverse demographic trend has been invoked as a further possible driver. To assess the role of demography two variables are used, the share of the population older than or equal to~65 (\textit{Age\,65 (\%)}) and the share of population older than or equal to~75 (\textit{Age\,75 (\%)}).

\subsection{Natural Resources}

The oil price is a well-known source of inflationary pressures in the world economy, and the change in the oil price is used as a control variable in several empirical studies. We consider two energy-related variables, both from the World Bank. The first is a weighted average of energy prices including coal, crude oil and gas (\textit{En. Prices (USD)}). The second is the total of natural resources rents (in \% of GDP). It is the sum of rents on oil, natural gas, coal, mineral, and forest, calculated as the difference between the price of a commodity and the average cost of producing it (\textit{En. Rents (\% GDP)}).

\subsection{Past Inflation}

In empirical studies past inflation is often controlled for. Countries that experienced high inflation might be more aware of its negative consequences and oppose it more forcefully. A related effect that can be assessed by past inflation rates is inflation inertia, according to which inflationary shocks may translate into higher inflation expectations through wage and price contracts, which in turn materialize in terms of higher actual inflation. We account for the influence of inflation in the past by constructing a 3-year moving geometric average of inflation (\textit{Past Infl. (\%)}).


\end{document}